\documentclass[useAMS,usedcolumn,usenatbib,usegraphicx,sort]{mn2e}
\usepackage{epsfig}
\usepackage{amssymb}
\usepackage{gensymb}
\usepackage{amsfonts}
\usepackage{amsmath}
\usepackage{color}
\usepackage{lscape,longtable}
\usepackage{times}
\usepackage{tablefootnote}
\usepackage{subfig}
\usepackage{caption}
\long\def\symbolfootnote[#1]#2{\begingroup%
\def\thefootnote{\fnsymbol{footnote}}\footnote[#1]{#2}\endgroup} 
\def\apjl{ApJ}%          % Astrophysical Journal, Letters
\def\apjs{ApJS}%          % Astrophysical Journal, Supplement
\def\aap{A\&A}%          % Astronomy and Astrophysics
\title[Eight new z $\geqslant$ 6 quasars ]
{Eight new luminous z $\boldmath \geq$ 6 quasars selected via SED model 
fitting of  
VISTA, WISE and Dark Energy Survey Year 1 Observations} 
\author[S.L. Reed, R.G. McMahon, P. ~Martini et al.]
{\parbox{\textwidth} 
{S. L. Reed$^{1,2}$\thanks{E-mail: sr525@ast.cam.ac.uk}, 
R. G. McMahon$^{1,2}$, P. ~Martini$^{4,5}$, M. ~Banerji$^{1,2}$, M.
~Auger$^{1}$, P.C. ~Hewett$^{1}$, S. E. ~Koposov$^{1}$, S. L. J. Gibbons$^{1}$, E.
Gonzalez-Solares$^{1}$, F. ~Ostrovski$^{1,2}$, S. S. ~Tie$^{4,5}$,
F.~B.~Abdalla$^{3,6}$, S.~Allam$^{7}$, A.~Benoit-L{\'e}vy$^{7,8,9}$,
E.~Bertin$^{8,9}$, D.~Brooks$^{3}$, E.~Buckley-Geer$^{7}$,
D.~L.~Burke$^{10,11}$, A. Carnero Rosell$^{12,13}$, M.~Carrasco~Kind$^{14,15}$, J.~Carretero$^{16,17}$, L.~N.~da Costa$^{12,13}$,
D.~L.~DePoy$^{18}$, S.~Desai$^{19}$, H.~T.~Diehl$^{7}$, P.~Doel$^{3}$,
A.~E.~Evrard$^{20,21}$, D.~A.~Finley$^{7}$, B.~Flaugher$^{7}$,
P.~Fosalba$^{22}$, J.~Frieman$^{7,23}$, J.~Garc\'ia-Bellido,
E.~Gaztanaga$^{22}$, D.~A.~Goldstein$^{24,25}$, D.~Gruen$^{26,11}$,
R.~A.~Gruendl$^{27,28}$, G.~Gutierrez$^{7}$, D.~J.~James$^{29,30}$,
K.~Kuehn$^{31}$, N.~Kuropatkin$^{7}$, O.~Lahav$^{3}$, M.~Lima$^{32,12}$,
M.~A.~G.~Maia$^{12,13}$, J.~L.~Marshall$^{18}$, P.~Melchior$^{33}$,
C.~J.~Miller$^{20,21}$, R.~Miquel$^{34,17}$, B.~Nord$^{7}$,
R.~Ogando$^{12,13}$, A.~A.~Plazas$^{35}$, A.~K.~Romer$^{36}$,
E.~Sanchez$^{37}$, V.~Scarpine$^{7}$, M.~Schubnell$^{21}$,
I.~Sevilla-Noarbe$^{37}$, R.~C.~Smith$^{30}$, F.~Sobreira$^{12,38}$,
E.~Suchyta$^{39}$, M.~E.~C.~Swanson$^{28}$, G.~Tarle$^{21}$,
D.~L.~Tucker$^{7}$, A.~R.~Walker$^{30}$, W.~Wester$^{7}$\\ Affliations at end
of paper.  }}

\begin{document}
\maketitle

\begin{abstract}
We present the discovery and spectroscopic confirmation with the ESO NTT and
Gemini South telescopes of eight new 
%and rediscovery of two previously known 
6.0 $<$ z $<$ 6.5 quasars with z$_{AB}$ $<$ 21.0.  These quasars were 
photometrically selected without any star-galaxy morphological criteria from
1533 deg$^{2}$ using SED model fitting to photometric data from the Dark
Energy Survey (g, r, i, z, Y), the VISTA Hemisphere Survey (J, H, K) and the
Wide-Field Infrared Survey Explorer (W1, W2). The photometric data was fitted
with a grid of quasar model SEDs with redshift dependent Lyman-$\alpha$ forest
absorption and a range of intrinsic reddening as well as a series of low mass
cool star models.  Candidates were ranked using on a SED-model based
$\chi^2$-statistic, which is extendable to other future imaging surveys (e.g.
LSST, Euclid).
%rather than conventional colour cut criteria. which allows more unusual objects 
%to be found such as VDESJ2250$-$5015 which has red colours (E(B-V)=0.1) and 
%would have be missed by previous surveys. 
Our spectral confirmation success rate is 100\% without the need for follow-up
photometric observations as used in other studies of this type.  Combined with
automatic removal of the main types of non-astrophysical contaminants the
method allows large data sets to be processed without human intervention and
without being over run by spurious false candidates. We also present a robust
parametric redshift estimating technique 
%based on the onset of the Lyman-alpha forest 
that gives compararable accuracy to MgII and CO based redshift estimators.  We
find two z $\sim$ 6.2 quasars with HII near zone sizes $\leq$ 3 proper Mpc which
could indicate that these quasars may be young with ages $\la 10^6 - 10^7$
years or lie in over dense regions of the IGM.  The z = 6.5 quasar
VDESJ0224$-$4711 has J$_{AB}$ = 19.75 is the second most luminous quasar known
with z $\geq$ 6.5.
\end{abstract}

\begin{keywords} dark ages, reionisation, first stars --- galaxies: active ---
galaxies: formation --- galaxies: high redshift -- quasars individual:
VDESJ0224$-$4711
\end{keywords}

\section{Introduction}

Quasars are some of the most luminous sources in the high redshift universe
and can be used as direct probes of very early times when the first generations
of galaxies and quasars were forming. Their spectra can be used to shine light
on the properties of the inter-galactic medium (IGM) as well as giving direct
measurements of the neutral hydrogen fraction at the end of reionisation
through the study of Ly$\alpha$ forest absorption (\citealt{Fan2006b,
Bolton2007}). Absorption lines in the spectra of high redshift quasars
allows the properties of gas and metals to be studied on cosmological scales.

The results from the Cosmic Microwave Background (CMB) measurements given in
\citet{Planck2015XIII} suggest that the beginning of reionisation was at
z $\sim$ 8. At lower redshifts ( 2.0 $<$ z $<$ 6.0) studies (\citealt{Gunn1965,
Fan2006b, Becker2007}) show that the IGM is highly ionised ($n_{\rm
{HI}}/n_{\rm H} \le 10^{-4}$) and therefore that reionisation was complete by
z $\sim$ 6. The discovery of more quasars above  a redshift of z = 6 will
allow the change in hydrogen ionisation at z$>$6 to be studied in 
more detail and along different lines of sight.

There have been many surveys for high redshift quasars and these have led to
the discovery of $\sim$ 60 (z $>$ 6.0) quasars (e.g. \citealt{Venemans2015,
Carnall2015, Willott2010, Jiang2009, Mortlock2012, Venemans2013, Fan2006b,
Banados2016, Jiang2016}).  Most of these searches have used purely optical
photometry from large surveys such as the Sloan Digital Sky Survey (SDSS) or
the Canada France Hawaii Telescope Legacy Survey (CFHTLS) which have a reddest
photometric waveband of z.  The deeper and redder photometry extending to the
Y photometric waveband provided by the Dark Energy Survey (DES)
\citep{Abbott2005} 
%means that accurate photometry can be obtained for
%sources at the average magnitude of high redshift quasars. This 
combined with the additional IR data from complementary surveys such as the
VISTA Hemisphere Survey (VHS) \citep{McMahon2013} and the Wide-field Infrared
Survey Explorer (\textit{WISE}) means that samples can be cleanly selected
without the need for deep photometric follow-up such as in \citet{Reed2015}.
Infrared data are a powerful discriminant between high redshift quasars and
their main astrophysical contaminants of ultra cool stars.
(\citealt{Wright2010, Banerji2015}).

The red sensitive Dark Energy Camera (DECam) CCD detectors, coupled with the
long wavelength sensitivity of the DES z and Y filters, allows the detection of
Ly-$\alpha$ to higher redshift than was possible with less red sensitive
optical surveys such as SDSS increases the redshift range that can be covered
to z$\sim$7. In this paper we present the results of our search for
high-redshift quasars in the first year of DES data.

DES magnitudes, near infrared (NIR) VISTA magnitudes and \textit{WISE}
magnitudes are quoted on the AB system.  The conversions from Vega to AB that
have been used for the VISTA data are: $J_{\textrm{AB}} = J_{\textrm{Vega}}
+ 0.937$ and $Ks_{\textrm{AB}} = Ks_{\textrm{Vega}} + 1.839$, these are taken
from the Cambridge Astronomical Survey Unit's website
\footnote{http://casu.ast.cam.ac.uk/surveys-projects/vista/technical/filter-set}.
The conversions used for the ALLWISE data are $W1_{\textrm{AB}}
= W1_{\textrm{Vega}} + 2.699$ and $W2_{\textrm{AB}} = W2_{\textrm{Vega}}
+ 3.339$ which are given in \citet{Jarrett2011} and in the \textit{ALLWISE}
explanatory supplement.~\footnote{The \textit{ALLWISE} explanatory supplement,
http://wise2.ipac.caltech.edu/docs/release/allwise/expsup/sec5\_3e.html,
directs the reader to the \textit{WISE All-Sky} explanatory supplement for the
conversions;
http://wise2.ipac.caltech.edu/docs/release/allsky/expsup/sec4\_4h.html\#summary.}
When required, a flat cosmology with $\Omega_{\mathrm{m0}} = $ 0.3 and
H$_{\mathrm{0}} = $ 70.0 km/s/Mpc was used. The code used in this analysis
makes use of the astropy python package \citep{AstColl}.

\section{Photometric Imaging Data}
\subsection{Dark Energy Survey Data}

We here use the  Year One First Annual (Y1A1) internal collaboration release of
the Dark Energy Survey (DES) data (\citet{Diehl2014}, Drlica-Wagner et al. (In
preparation)). These data cover $\sim$1840 deg$^{2}$ of the southern celestial
hemisphere to a median 10$\sigma$ point source (MAG\_PSF) depth in AB
magnitudes of 23.28, 23.6, 23.1, 22.3 and 20.8.  in the g, r, i, z and Y bands
respectively.  Catalogue source detection uses the SExtractor
\citep{Bertin1996} image detection software in double image mode using the
$\chi^2$ detection image \citep{Szalay1999} constructed form the
combination of the r, i and z band images, as the detection image.

\begin{figure} \includegraphics[width = \linewidth]{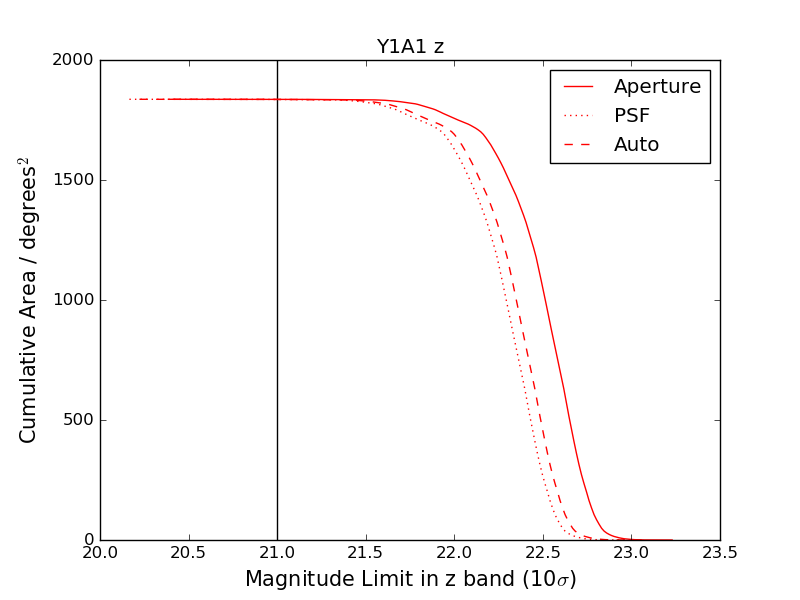}
\caption{Cumulative area versus 10$\sigma$ $z$-band depth 
in a 2$\arcsec$ diameter aperture (solid line), PSF magnitudes (dotted line)
and auto magnitudes (dashed line) for the DES Y1A1 data. The aperture
magnitudes were converted to PSF and auto magnitudes using the median offset
between PSF (or auto) magnitude and aperture magnitude for point sources from
the whole Y1A1 dataset.  Our magnitude limit of $z_{PSF} < 21.0$ is shown as
the vertical line. 
Auto magnitudes are intended to give the most
precise estimate of total magnitudes for galaxies. We do not use
auto magnitudes in the analysis and they are included here for comparison with
other work.}
\label{zMagLim} \end{figure}

The point source depths are calculated from area weighted median aperture
(MAG\_APER\_4) magnitude limits taken from the DES Mangle \citep{Swanson2008}
products at 10$\sigma$ in a 2" diameter aperture values of 24.2, 23.9, 23.3,
22.5 and 21.2. See Figure \ref{zMagLim} for the z band magnitude limit versus
cumulative area.  To convert these aperture depths to point source deths,
the median differences between the 2" aperture magnitude and the PSF magnitude
for point sources were used across the survey. The offsets (MAG\_APER $-$
MAG\_PSF) are g: -0.4, r: -0.3, i: -0.2, z: -0.2, Y: -0.4 with the differences
due to the differences in the average point spread function widths induced by
seeing for each waveband.

Once completed DES will cover 5000 deg$^2$ in five optical bands with images
taken using the Dark Energy Camera (DECam) \citep{Flaugher2015} which is
mounted on the Blanco 4-meter telescope at the Cerro Tololo Inter-American
Observatory (CTIO). The data are then reduced using the DES data management
process (\citealt{Mohr2012, Desai2012}).  DECam is particularly suited to high
redshift survey work due to its large field of view and red sensitivity. The
data contained in the Y1A1 release were taken between 2013 August 15$^{\textrm
{th}}$ and 2014 February 9$^{\textrm {th}}$. The Y1A1 release is shallower than
the final survey depth and consists of 3707 coadded tiles covering two
contiguous regions one overlapping the Stripe 82 area imaged by the SDSS and
one overlapping with the area covered by the South Pole Telescope (SPT).  The
tiles are coadd images made up of between 1 and 5 exposures in each of the
5 wavebands with an average coverage of 3.5 exposures making up each tile.

A magnitude limit of z$_{\mathrm{PSF}}$ $\leq$ 21.0 was used in this work; this
corresponds to an area of 1835 deg$^{2}$. Figure \ref{zMagLim} shows the
cumulative area against depth for the dataset in 2" diameter aperture, PSF and
auto magnitudes for stellar objects where we define stellar objects based on
\citet{Reed2015}.  Whilst auto magnitudes are intended to give the most precise
estimate of total magnitudes for galaxies they can also be used for stellar
objects. The SExtractor implementation of the routine is based on
\citet{Kron1980}. We do not use auto magnitudes in the analysis here and they
are included here for comparison with other work noting the small offset for
point sources between auto magnitudes and PSF magnitudes which may indicate
a systematic over estimate in the auto fluxes. The z band limit used here is
shown as the vertical line and is well above the 10$\sigma$ limit.  As we only
used the area of DES currently covered with VHS ($\sim$84\%) this reduced
the total area available to 1533 deg$^{2}$.

In this paper DES magnitudes quoted are point spread function (PSF) magnitudes
derived from PSF fluxes calculated using PSFs for each coadd tile
measured as part of the DES reduction using PSFex \citep{Bertin2011}. When
other magnitude or flux measurements (e.g. aperture) are used this is
explicitly stated.  All magnitudes are given in the AB system.  Aperture
magnitudes and fluxes from DES are given for a 2" diameter aperture with an
aperture correction applied based on the point spread function to compensate
for missing flux outside the aperture unless otherwise stated.  Corrected
aperture fluxes were used in the model fitting calculations and the fluxes
given in the paper are aperture corrected fluxes unless otherwise stated.
Aperture flux measurement were used as they best represent the flux when the
object was near or below the detection limit of the data.

\subsection{VISTA Hemisphere Survey Data}
The VISTA Hemisphere Survey (VHS) \citep{McMahon2013} aims to carry out a near
infra red (NIR) survey of $\sim$ 18,000 degrees$^{2}$ of the southern
hemisphere to a depth 30 times fainter than the Two Micron All Sky Survey
(2MASS) in two wavebands J and K$_{s}$. The survey uses the 4m VISTA telescope
at ESO's Cerro Paranal Observatory in Chile.  In the Southern Galactic Cap
$\sim$5000 degrees$^{2}$, which will overlap the DES area, is being imaged more
deeply (J$_{AB}$ = 21.2, K$_{s;AB}$ = 20.4; 5$\sigma$ point source depths) with
partial coverage in H.  This gives data in three bands (J, H and K$_{s}$) in
the near infrared at $\sim$1-2$\mu$m. H band data is not being taken over the
full DES and some of the area used in this project does not have H band
imaging. The VHS data used in this work were taken between
2009 November 4$^{\textrm {th}}$ and 2014 February 1$^{\textrm {st}}$.

The VIRCAM camera \citep{Dalton2006} used for VHS imaging has a sparse array of
16 individual 2k x 2k MCT detectors covering a region of 0.595 square degrees.
In order to cover the full 1.5 square degree field of view of the camera six
exposures are required. These exposures are then combined into one co-added
tile as part of the pipeline processing. The data is processed with the VISTA
Data Flow System at CASU (\citealt{Irwin2004, Emerson2004, Lewis2010}) and the
science products are available from the ESO Science Archive Facility and the
VISTA Science Archive (\citealt{Hambly2004, Cross2012}).

\subsection{Wide Infrared Survey Explorer Data}
Longer wavelength data at 3.4, 4.6, 12 and 22$\mathrm\mu$m (known as W1, W2, W3
and W4 respectively) were used from the all-sky Wide Infrared Survey Explorer
(\textit{WISE}) \citep{Wright2010}. The \textit{WISE} satellite uses a 40cm
telescope with a camera containing four 1024 x 1024 arrays with a median 
pixel size of 2.757" and a field of view of 47 x 47 arcminutes. The telescope
scanned the sky and took multiple images 
%with an individual exposure time of 7.7 seconds in W1 and W2 and 8.8 seconds
%in W3 and W4 
giving coadd 5$\sigma$ point source depths of W1$_{\textrm{AB}}$ = 19.3,
W2$_{\textrm{AB}}$ = 18.9, W3$_{\textrm{AB}}$ = 16.5 and W4$_{\textrm{AB}}$
= 14.6. In W1, W2 and W3 these coadd images had a full width at half maximum of
6.1" and in W4 6.4". Once the cryogenic fuel was exhausted in 2010 the
telescope continued to survey the sky in the two shortest bands as part of the
post-cryogenic \textit{NEOWISE} mission phase. The two datasets were combined
into the 2013 \textit{WISE} AllWISE Data Release.  The AllWISE coadd images are
4095 $\times$ 4095 pixels at 1.375 arcsec per pixel.

\section{Quasar Candidate Selection} \label{SelectionCriteria}

\begin{figure} 
\includegraphics[width = \linewidth]{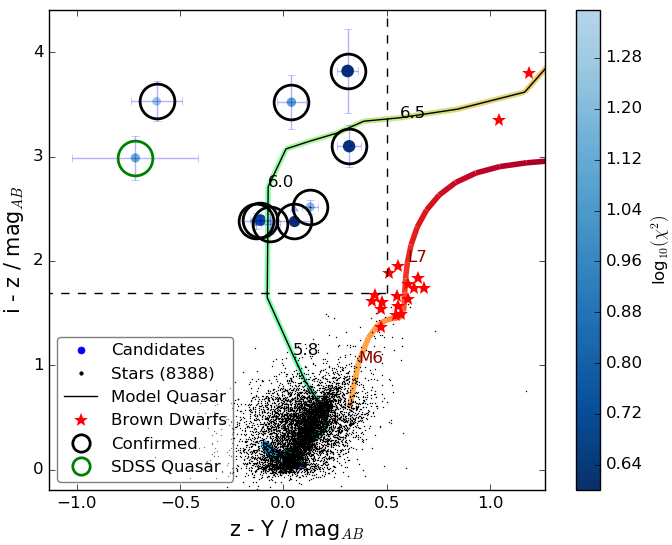} 
\caption{z - Y versus i - z colour - colour
diagram shows the colour space used for the selection. The dashed lines show
the z - Y and i - z colour cut limits used. This colour cut limit is the same
as was used in R15 and was designed to help remove cool stars.
% In future work we hope to loosen or remove this colour cut to ensure 
% we are recovering the reddest quasars. 
The black points are stars taken from three tiles of DES data and the
red stars are known brown dwarfs from \citet{Kirkpatrick2011} matched to the
DES data. The red line shows the derived colour track for dwarf stars; the
colour of the line corresponds to the colour of the line in figures \ref{QFit}
and \ref{BDFit} as does the colour of the blue-green line which shows one of
the quasar tracks used. The blue points give our candidate objects with higher
ranked objects being darker in colour and larger in size. The black circled
objects were followed up spectroscopically, the green circle shows the known
SDSS quasar. Objects with a good fit to the brown dwarf model are not shown on
this plot. A large colour region around the predicted colour line is
probed to account for the intrinsic variation in the SEDs of quasars as well as
line of sight extinction in the sources and the uncertainties in the photometry
for each object.} \label{izzY} 
\end{figure}

\begin{figure}
\includegraphics[width = \linewidth]{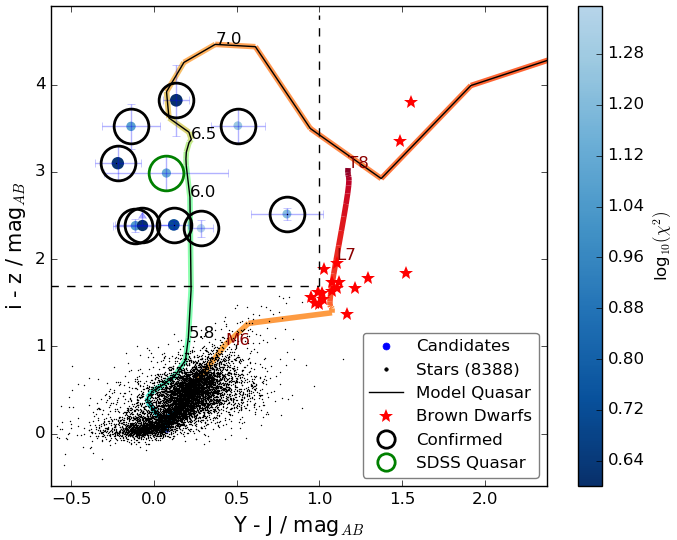}
\caption{Y - J versus i - z colour - colour plot showing the reasoning
behind the Y - J colour cut. A colour limit of 1.0 (marked with a dashed line)
removes all known coll dwarf stars while allowing us to probe as much parameter
space as possible. In future work we hope to loosen or remove this colour cut.
The points in this Figure follow the same schema as in Figure \ref{izzY}.
Objects with a good fit to the brown dwarf model are not shown on this plot.
} \label{izYJ} \end{figure}

Following on from the selection method presented in \citet{Reed2015} (hereafter
R15) we have developed a selection method that uses all the photometric
data (from \textit{WISE}, VHS and DES) available for the objects. The selection
method incorporated the first eight steps outlined in section 3 of R15 and is
summarised in Table \ref{tab:selection}. Then the candidate list
was matched to the VHS catalogue data to give J and K band magnitudes for the
objects and was a fast way to remove artefacts such as cosmic rays that were
present in only one of the surveys. Matching to VHS and keeping only objects
with Y$_{AB}$ - J$_{AB}$ $<$ 1.0 left 960 candidates from the original 4,195
that satisfied the first stages of selection. The cuts used are shown in Figure
\ref{izzY} and \ref{izYJ}. The z - Y and Y - J cuts were chosen to be the
reddest cuts that excluded all known dwarf stars used in this analysis.

%\begin{figure} 
%\includegraphics[width = \linewidth]{iz_zY_clean.png}
%\caption{A colour - colour plot showing the colour cuts used to cut down the
%initial input list. The dashed lines show our colour cuts, the black points are
%stars taken from three tiles of DES data and the red stars are known brown
%dwarfs matched to the DES data. The red line shows the derived colour track for
%dwarf stars; the colour of the line corresponds to the colour of the line in
%figures \ref{QFit} and \ref{BDFit} as does the colour of the blue line which
%shows one of the quasar tracks used. The blue points give our candidate objects
%with higher ranked objects being darker in colour and larger in size. The black
%circled objects were followed up spectroscopically, the green circle shows the
%known SDSS quasar and the grey circles show the two candidates that were ranked
%higher than some of the objects we observed but that we did not prioritise for
%follow-up. Objects with a good fit to the brown dwarf model are not shown on
%this plot. } \label{izzY} \end{figure}

\subsection{\textit{WISE} List Driven Aperture Photometry}
As the W1 - W2 colour is a discriminant between quasars and cool
stars, list driven aperture photometry code was run on the unWISE images
\citep{Lang2014}. We use these unblurred coadds from the \textit{WISE} Atlas
images.  The unWISE coadd images are 2048 $\times$ 2048 pixels at the nominal
native pixel scale, 2.75 per arcsec pixel, rather than the 4095 $\times$ 
4095 images at 1.375 per pixel chosen in the AllWISE Release.  The blurred
AllWISE images are better suited for source detection but the unblurred unWISE
coadds used here have better resolution, and are therefore more appropriate for
forced, list-driven photometry at known positions on the sky.
\citep{Lang2014}.  Forced photometry was done using imcore\_list from version
1.0.26 of the CASUTOOLS software package \citep{Irwin1985}
\footnote{http://casu.ast.cam.ac.uk/surveys-projects/software-release}. The
code was run for all objects in the VHS catalogue where the VHS sources were
used as inputs as they are closer in wavelength than the DES sources so were
more likely to have a corresponding \textit{WISE} source. The approach of using
the J band positions will also allow us to use these same catalogues as we push
our search to higher redshift where we expect quasars candidates to longer
detected in the shorter (riz) wavelength DES bands at all.  An aperture radius
of 2.5 pixels ($\sim$ 6.9") was used for the list driven photometry on the WISE
images. This was chosen to match the aperture size used for the published
\textit{WISE} catalogues.  While a smaller aperture size would help to ensure
that the flux came only from the specified object it would also miss more of
the WISE flux that is outside the aperture due to the point spread function of
\textit{WISE}. A larger aperture can also include flux from neighbouring
objects. An alternative approach would be to estimate the WISE fluxes using PSF
based weighting.
%{\bf See Section 999 for a further discussion}.
% In future versions of the forced photometry
%catalogue we hope to more accurately assign flux to its originating object.
\begin{table} \begin{center} 
\caption{Summary of the steps in the high-redshift quasar selection process.
The individual parts of step one are detailed fully in R15 and are not
differentiated here.} \label{tab:selection} 
\begin{tabular}{cccc}
\hline Step & Description & Number& Number\\ &  & Removed & Remaining \\ \hline
& Number of objects in database & & 139,142,161  \\
1 & Steps 1-8b from R15$^{\dagger}$ & & \\
 & z$_{\mathrm{PSF}} \leq$ 21.0 and $\sigma_{\mathrm{z}} <$ 0.1 & &  \\
 &  i$_{\mathrm{PSF}}$ - z$_{\mathrm{PSF}} <$ 1.694 &  & \\
 & g$_{\mathrm{PSF}}$ and r$_{\mathrm{PSF}} >$ 23.0 & & \\
 & $\sigma_{\mathrm{g}}$ and $\sigma_{\mathrm{r}} >$ 0.1 & & \\
 & z$_{\mathrm{PSF}}$ - Y$_{\mathrm{PSF}} <$ 0.5 & & \\
 & Y$_{\mathrm{PSF}} <$ 23.0 & 139,135,538 & 4,195 \\
2 & Y - J $<$ 1.0 & 3,235  & 960 \\
3 & Remove Chip Edges in z Band & 498 & 462 \\
4 & Remove Bad Image Areas & 105 & 393 \\
5 & Remove Objects Bright in r & 246 & 147 \\
\hline
\vspace{-8pt}\\
\multicolumn{4}{l}{\scriptsize{$^{\dagger}$A magnitude limit of z = 21 was used
rather than 21.5 in R15 and no point source separation.}}\\ \end{tabular}
\end{center}
\end{table}

\subsection{Photometric SED modelling, redshifts and stellar classification}
\label{Section:PhotoZ}
To prioritise the candidates a photometric redshift fit was carried out using
a series of model spectral energy distributions (SEDs).  Four quasar models
\citep{Maddox2012} based on the spectral templates in \citet{Maddox2006}, with
different levels of intrinsic reddening ($E(B-V) = 0.0, 0.025, 0.05, 0.10$) 
were used in 0.1 redshift increments between 4.0 and 8.0 for the model
fitting.  The model is a parametric model where the continuum consists of two
power laws (with slopes -0.42 and -0.17) that are joined at 2340\textrm{\AA}.
Longward of one micron the flux is dominated by a single temperature black body
with T = 1236K. On top of this is an empirical quasar emission line spectrum.
Shortward of the Ly$\alpha$ emission line the continuum flux is suppressed by
a model the Ly$\alpha$ forest absorption which is redshift dependent. All the
flux shortward of the restframe Lyman-limit (912\textrm{\AA}) is removed. Thus
at all redshifts above z = 5; there should be zero flux in the DES g band which
has $<$1\% of peak transmission at $\lambda > 5530\textrm{\AA}$.  The flux from
the model was integrated over all the DES and VHS wavebands as well as the
\textit{WISE} W1 and W2 bands. As the DES aperture fluxes do not include
aperture corrections by default and SExtractor \citep{Bertin1996} does not
return negative fluxes for PSF fluxes, aperture corrections were calculated to
account for any flux that fell outside the aperture. It was necessary to have
good measurements of the flux for very faint/undetected objects as all of our
candidates are not present in the bluest DES bands. The aperture corrections
were calculated using the median of the $\frac{PSF flux}{Aperture flux}$ for
stellar objects. They were calculated for each individual DES image tile and
applied separately for each tile. The objects were also compared to the derived
brown dwarf colours from \citet{Skrzypek2015}. As these colours were given in
the UKIRT Infrared Deep Sky Survey (UKIDSS) Large Area Survey (LAS) and SDSS
pass bands, colour terms (these are given in \ref{CTerms}) were calculated
between the surveys using the overlap between DES, UKIDSS, VHS and SDSS in
Stripe 82. The colours were then converted onto the AB system using the offsets
given in \citet{Hewett2006}. Table \ref{tab:QProps} shows the ten objects
followed up in this work and Table \ref{tab:BDs} shows the ten objects ranked
most highly to be brown dwarfs.  Figures \ref{QFit} and \ref{BDFit} show the
results of the model fitting for the highest ranked quasar candidate and a 
probable low mass star with spectral type M7.

The reduced $\chi^2$ ($\chi^2_{\mathrm{reduced}}$) values were derived using the formula 
below:

\begin{equation}
\label{equation::chi2_a}
\chi^2_{\mathrm{i,reduced}} = \sum_{n=1}^N \left(\frac{\mathrm{data}_n - f_n(\mathrm{model_i)}}{\sigma(\mathrm{data})_n}\right)^2 \bigg/ (N-1)
\end{equation}

where for each $\rm \rm model_i$ we sum over n = 1...N wavebands 
with N-1 degrees of freedom.

\begin{figure*}
\includegraphics[width = \linewidth]{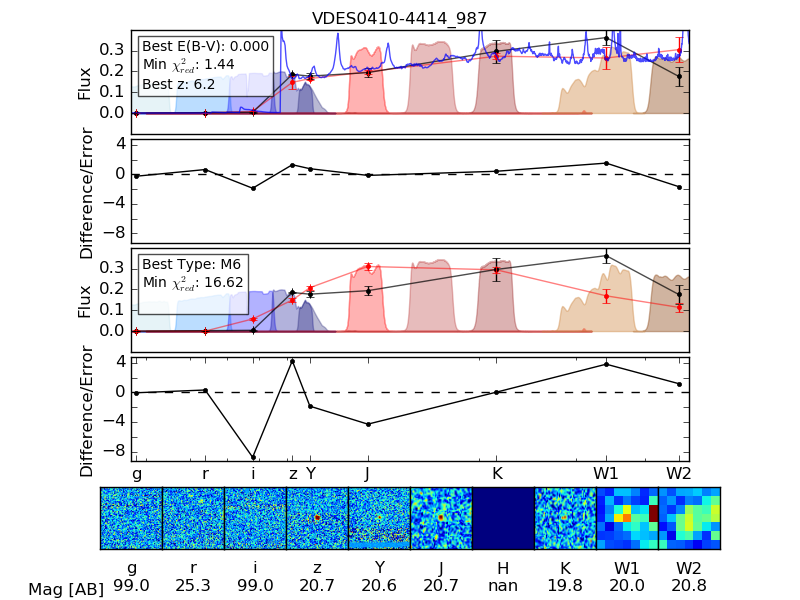}
\caption{An example of the model fitting results for the highest ranked quasar in the
sample. The top panel shows the best fitting quasar model in red and the data
with associated uncertainties in black. The filled areas show the filters for DES, VHS
and \textit{WISE}. The blue line shows the model quasar spectra used. The
second panel down shows the residuals from the fit divided by the uncertainties on each
point. The bottom two panels show the same thing for the best fitting cool star
model. Along the bottom of the Figure are 20" cutouts in each band with the AB
magnitude in that band beneath.}
\label{QFit}
\end{figure*}

\begin{figure*}
\includegraphics[width = \linewidth]{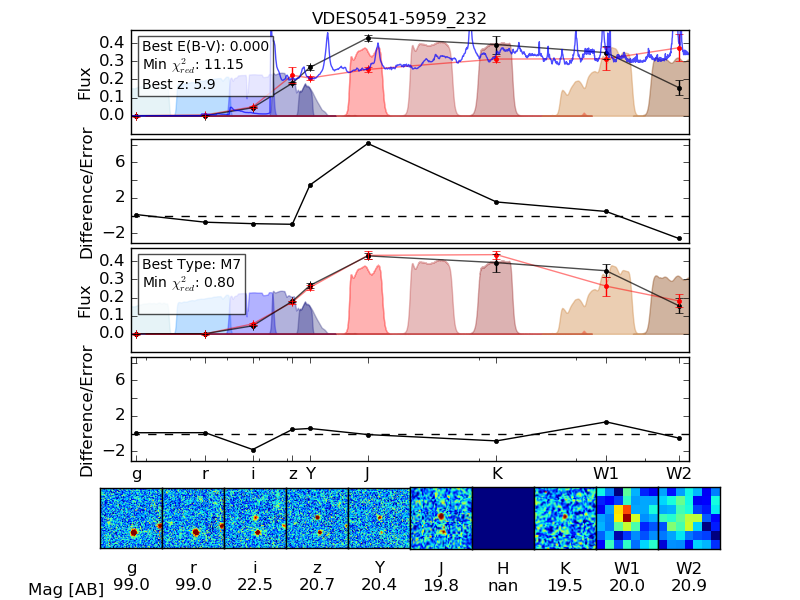}
\caption{An example of the fitting results for the highest ranked brown dwarf in the
sample. The colours and lines are the same as in Figure \ref{QFit}.}

\label{BDFit}
\end{figure*}

When the photometric fitting method was first run it was found that objects
with unreliable non-gaussian errors in their photometry, due for example to 
CCD chip edges and saturated objects, were contaminating the candidate list.
These objects were then removed using image based techniques. To remove objects
with photometry effected by chip edges the pixel values in a 30" box around the
object were analyzed and if more than a third of had the same value the object
was rejected; this also removes areas which have been masked with zeros in the
image (such as saturated areas and bleed trails). It was found that a large
number of the candidates appeared to have no measured flux in the g, r, or
i bands but there were also no other objects present with a region with radius
of 30''  around the location of the candidate in the image . It was found that
these patches of image had very different noise properties compared to other
parts of the image. To remove these the median and the median absolute
deviation (MAD) of the pixel values in a 30" box around the object were
calculated and objects with a MAD $<$ 0.7 were removed. This pixel MAD
threshold was derived empirically from studying a large number of images. The
distribution of MAD values of boxes taken from across a range of images was
found to be bimodal with a dip at $\sim$ 0.7.  The bright r band objects were
removed as detailed in R15.  These pixel level filtering steps are done after
catalogue level selection as the image based techniques are more
computationally intensive than the catalogue ones so it is more efficient to
run them on the reduced candidate list rather than to create a completely clean
list from the beginning. Furthermore, if the images are not available colocated
to computational resources, network transfers can be prohibitive. Table
\ref{tab:selection} lists the numbers of candiates removed by each selection
stage.

The photometric fitting was then run again on the 147 remaining candidates.
Candidates were first ranked based only on their quasar reduced $\chi^{2}$
values with the smallest reduced $\chi^{2}$ sources having the highest ranking.
Following this ranking, we visually inspected the candidates in ranked order to
remove artefacts and junk sources, and also compared the quasar reduced
$\chi^{2}$ values to those obtained from a brown dwarf fit to the photometry.
The likelihood of being a brown dwarf was calculated from the polynomial fits
in \citet{Skrzypek2015}. Objects where the reduced $\chi^{2}$ to be a brown
dwarf was comparable to or higher than that to be a quasar were removed.

We found that the reduced $\chi^{2}$ values for the best fitting quasar and low
mass star models often exceeded 3 and hence were ruled out at $>99\%$. At face
value this is indicative that neither model fitted the data. This could be
interpreted to mean that the photometric measurements had systematic errors or
the range of SED models being considered was not representative of the
underlying true distribution.  We took a pragmatic approach and added
a systematic photometric uncertainty to the statistical uncertainty in each
waveband. Percentage errors in flux of 10\%, 10\%, 10\%, 20\%, 5\%, 5\%, 5\%,
5\%, 20\% and 20\% in g, r, i, z, Y, J, H, Ks, W1 and W2 respectively were
added in quadrature to the statistical uncertainties as show in Equation \ref{equation:chi2_b}

\begin{equation}
\label{equation:chi2_b}
\chi^2_{\mathrm{i,reduced}} = \sum_{n=1}^N \left(\frac{\mathrm{data}_n - f_n(\mathrm{model_i})}{\sigma(\mathrm{data+model})_n}\right)^2 \bigg/ (N-1)
\end{equation}

The resultant $\chi^{2}$
values for the 10 highest ranked most probable quasars are shown in Table
\ref{tab:QProps}. The 10 objects with highest low mass star SED probability are
listed in Table \ref{tab:BDs} and have a range of best fit spectral types from
M5 to L3.

\begin{table*} \begin{center} 
\caption{Parameters from the fitting process for the confirmed quasars.}
\label{tab:QProps} 
\begin{tabular}{lcccccccc}
\hline Name & Rank & $\chi^{2}_{red}$ of Best & Best & Best & Spectroscopic & $\chi^{2}_{red}$ of Best & Best & $\frac{\chi^{2}_{Q}}{\chi^{2}_{BD}}$ \\ 
&& Quasar Model & E(B-V) & Redshift & Redshift & Brown Dwarf Model & Type & \\ 
\hline

VDESJ0143-5545 & 9 & 3.15 & 0.100 & 6.1 & 6.25 & 38.87 & M7 & 0.081 \\
VDESJ0224-4711 & 3 & 1.62 & 0.050 & 6.4 & 6.50 & 32.24 & M7 & 0.050 \\
VDESJ0323-4701 & 10 & 3.35 & 0.000 & 6.1 & 6.25 & 15.02 & M5 & 0.223 \\
VDESJ0330-4025 & 5 & 2.24 & 0.025 & 6.2 & 6.25 & 18.71 & M7 & 0.120 \\
VDESJ0408-5632 & 8 & 3.10 & 0.000 & 6.0 & 6.03 & 13.76 & M6 & 0.225 \\
VDESJ0410-4414 & 1 & 1.44 & 0.000 & 6.2 & 6.21 & 16.62 & M6 & 0.087 \\ 
VDESJ0420-4453 & 6 & 2.54 & 0.000 & 6.0 & 6.07 & 19.44 & M6 & 0.131 \\
VDESJ0454-4448$^{\dagger}$ & 2 & 1.55 & 0.000 & 6.0 & 6.10 & 18.81 & M6 & 0.082 \\
VDESJ2250-5015 & 4 & 1.78 & 0.050 & 6.0 & 6.00 & 12.20 & M8 & 0.146\\
VDESJ2315-0023$^{\star}$ & 7 & 2.67 & 0.000 & 6.0 & 6.12 & 30.92 & M5 & 0.086 \\

\hline
\vspace{-8pt}\\
\multicolumn{4}{l}{\scriptsize{$^{\dagger}$This object was found in
R15.}}\\ 
\multicolumn{4}{l}{\scriptsize{$^{\star}$This object is SDSS
J231546.57+002358.1 found in \citet{Jiang2008}.}}\\
\end{tabular}
\end{center}
\end{table*}

\begin{table*} \begin{center} 
\caption{Parameters from the fitting process for the ten objects ranked highest
to be brown dwarfs. Only objects with photometry in all the available bands
were included here.} 
\label{tab:BDs} 
\begin{tabular}{lccccccc}
\hline Name & $\chi^{2}_{red}$ of Best & Best & Best & $\chi^{2}_{red}$ of Best & Best & $\frac{\chi^{2}_{Q}}{\chi^{2}_{BD}}$ \\ 
& Quasar Model & E(B-V) & Redshift & Brown Dwarf Model & Type & \\ 
\hline

VDESJ0419-5033 & 18.55 & 0.000 & 6.0 & 1.05 & L0 & 17.67\\
VDESJ0440-5258 & 11.48 & 0.000 & 6.0 & 1.43 & L0 &  8.03\\
VDESJ0516-5433 &  9.25 & 0.050 & 6.0 & 1.38 & L3 &  4.53\\
VDESJ0524-5710 & 19.37 & 0.000 & 5.9 & 0.82 & M7 & 23.62\\ 
VDESJ0541-5959 & 11.15 & 0.000 & 5.9 & 0.80 & M7 & 13.94\\ 
VDESJ2138-5853 & 18.69 & 0.000 & 5.9 & 1.36 & M7 & 13.75\\
VDESJ2248-4639 & 12.36 & 0.025 & 6.0 & 1.31 & L1 &  9.44\\
VDESJ2300-4432 & 13.88 & 0.000 & 6.0 & 1.06 & M9 & 13.10\\
VDESJ2307-0044 & 24.54 & 0.000 & 6.0 & 1.07 & L0 & 22.93\\
VDESJ2321-5655 &  6.66 & 0.000 & 5.2 & 1.10 & M5 &  6.05\\

\hline
\end{tabular}
\end{center}
\end{table*}

\section{Spectroscopic Observations}

\begin{table*}
\begin{center}
\caption{Details of the spectroscopic observations}
\label{tab:specObs}
\begin{tabular}{ccccccc}
\hline
Name & Telescope & Instrument & Exposure Time & Date & Filter & Grating/\\
 &  &  & (Seconds) &  & &Grism \\
\hline
VDESJ0143-5545 & NTT & EFOSC2 & 1200 + 1200 = 2400 & 09/11/2015 & OG530 & Gr\#16 \\
VDESJ0224-4711 & NTT & EFOSC2 & 1800 + 1800 = 3600 & 07/11/2015 & OG530 & Gr\#16 \\
VDESJ0323-4701 & GEMINI-SOUTH & GMOS-S & 600 + 600 + 600 + 600 = 2400 & 22/11/2015 & RG610\_G0331 & R400+\_G5325\\
VDESJ0330-4025 & GEMINI-SOUTH & GMOS-S & 600 + 600 + 600 + 600 = 2400 & 22/11/2015 & RG610\_G0331 & R400+\_G5325\\
VDESJ0408-5632 & NTT & EFOSC2 & 1200 + 1200 = 2400 & 08/11/2015 & OG530 & Gr\#16 \\
VDESJ0410-4414 & GEMINI-SOUTH & GMOS-S & 600 + 600 + 600 + 600 = 2400 & 11/09/2015 & RG610\_G0331 & R400+\_G5325\\ 
VDESJ0420-4453 & GEMINI-SOUTH & GMOS-S & 600 + 600 + 600 + 600 = 2400 & 04/09/2015 & RG610\_G0331 & R400+\_G5325\\
VDESJ2250-5015 & NTT & EFOSC2 & 1800 + 1800 = 3600 & 07/11/2015 & OG530 & Gr\#16 \\
\hline
\end{tabular}
\end{center}
\end{table*}

\begin{figure*} 
\includegraphics[width = \linewidth]{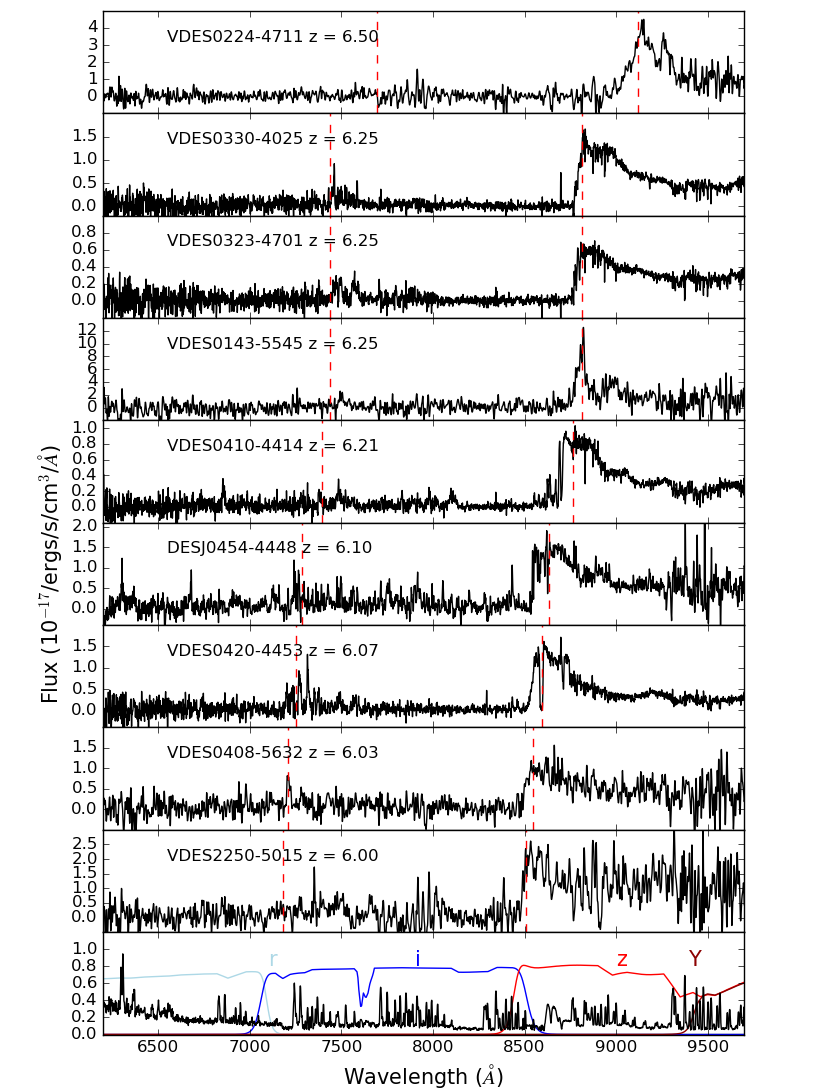}
\caption{Reduced spectra of all the objects in this sample as well as the
quasar discovered in R15 (DESJ0454-4448); presented in redshift order. The
vertical lines show the positions of Ly$\alpha$ and Ly$\beta$. The
bottom plot gives an example error spectra taken from one of the quasars
(DESJ0410-4414) and has the DES filters over plotted.}
\label{ManySpec}
\end{figure*}

Spectroscopic observations were obtained between 2015 October and November
using the European Southern Observatory's (ESO) 3.6m New Technology Telescope
(NTT) and the 8.1m Gemini-South Telescope. The confirmed quasars from these
spectroscopic follow up runs are listed in Table \ref{tab:QProps}.  A summary
of the observations, including the exposure times and grism/grating used, is
given in Table \ref{tab:specObs} and a summary of the objects' properties is
given in Table \ref{tab:properties}.  Figure \ref{ManySpec} shows the spectra
of the objects presented here along with the spectrum of the object detailed in
R15 as it was rediscovered in this sample. Four of the objects were observed
with the NTT at ESO's La Silla observatory over three nights from the 2015
October 7th to the 9th. The spectra were taken with the ESO Faint Object
Spectrograph and Camera 2 (EFOSC2) \citep{Buzzoni1984} and reduced using
a custom set of python routines.  Calibration data were taken during the
afternoon preceding the observations or taken as part of the PESSTO project
\citep{Smartt2015}. A 1.5" width slit was used and the data was binned 2x2 on
data readout.  Due to the inclemency of the weather due to partial cloud
coverage and the smaller mirror aperture the NTT data are of modest quality
compared to the Gemini observations of the rest of the  sample. Four of the
objects were observed with the Gemini Multi-Object Spectrograph (GMOS)
\citep{Hook2004} at the Gemini South Telescope as part of the 2015B queue
observations using a 0.75" mask and reduced with a custom python reduction
code.  All the reduced spectra are shown in Figure \ref{ManySpec}. A pipeline
was written to reduce the two dimensional spectra from both telescopes. The
object was located on the CCD and a gaussian was fitted to a well behaved area
of the spectrum. Standard star observations were used to study the change of
position of the spectrum in the spatial direction with wavelength. This was
found to vary little with time and a general formula for the trace was derived.
This could not be done from the quasar spectrum as it only covered a small
wavelength range at the reddest end of the detector. To be sure that we were
seeing no flux due to the intrinsic properties of the object rather than
because we were extracting the wrong part of the two dimensional spectrum this
trace was positioned using the small area of spectrum we have. The derived
gaussian profile was then used to weight the spectrum extracted along the line
of the trace.  Once the spectrum had been extracted the response function of
the instrument was calculated using the standard star observations and the
spectrum corrected.  The different spectrum were then stacked together and the
result was calibrated using the multi band photometry.  Wavelength calibration
was applied using arc lamp observations taken in the day prior to the
observations.

\subsection{Redshift Determination}
Redshifts were calculated by fitting a quasar model to the spectroscopic
data. The section of the spectra blueward of Ly$\alpha$ was modelled using an
exponential to account for the rapid decay to zero flux. A Gaussian centred at
1025.7$\mathrm{\AA}$ was used to approximate the Ly$\beta$ emission feature
seen in some of the spectra. Ly$\alpha$ emission was modelled using half
a Gaussian which matched onto the exponential at 1215.67$\mathrm{\AA}$.
Redward of Ly$\alpha$ the N{\sc{v}}, O{\sc{I}} and Si{\sc{IV}}+O{\sc{IV}}
lines were added using Gaussians centered at 1240.1, 1304.46 and 1397.8
$\mathrm{\AA}$ respectively \citep{Tytler1992}. The section longward of
1215.67$\mathrm{\AA}$ then had a power law and a constant offset added to model
the continuum emission. 

\begin{figure}
\includegraphics[width = \linewidth]{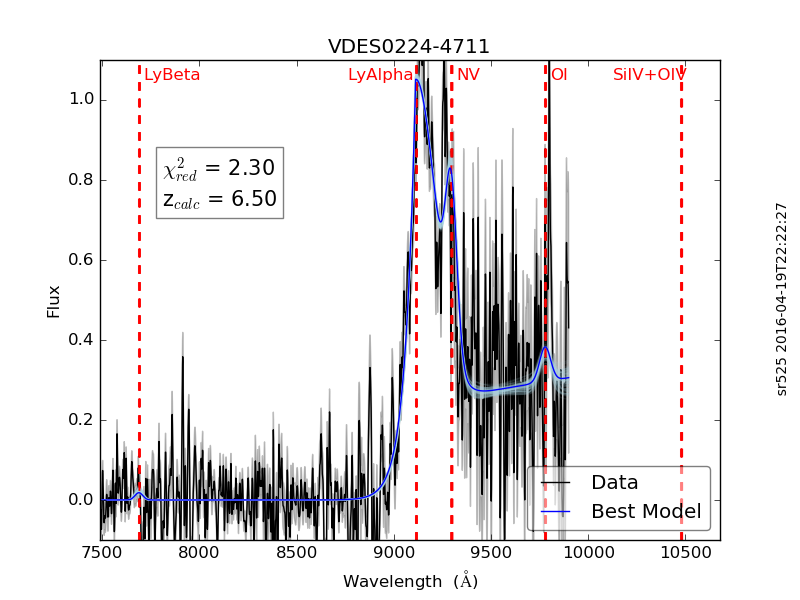}
\caption{An model fit for the highest redshift quasar in this sample. The
dashed lines show the centres of the lines used in the model. 
The data shown in
black is the unsmoothed spectrum and the grey shaded area shows the 
uncertainty at
each wavelength. The dark blue line is the best fitting model and the 
light blue lines show 100 example model fits found during the fitting 
iterations. The
reduced $\chi^2$ from the fit and the calculated redshift are given 
in the
inset panel.}
\label{SpecFitExample}
\end{figure}

This model was tested using the spectroscopic data from
\citet{Fan2006b}. While the spectroscopic data presented here do not cover the
full range of lines input into the model some of the test data covered the full
range. This model was then fitted to the data using a $\chi^{2}$ minimisation
to give the best estimate of the redshift. An example of the redshift fitting
process is shown in Figure \ref{SpecFitExample}.

\begin{figure} 
\includegraphics[width = \linewidth]{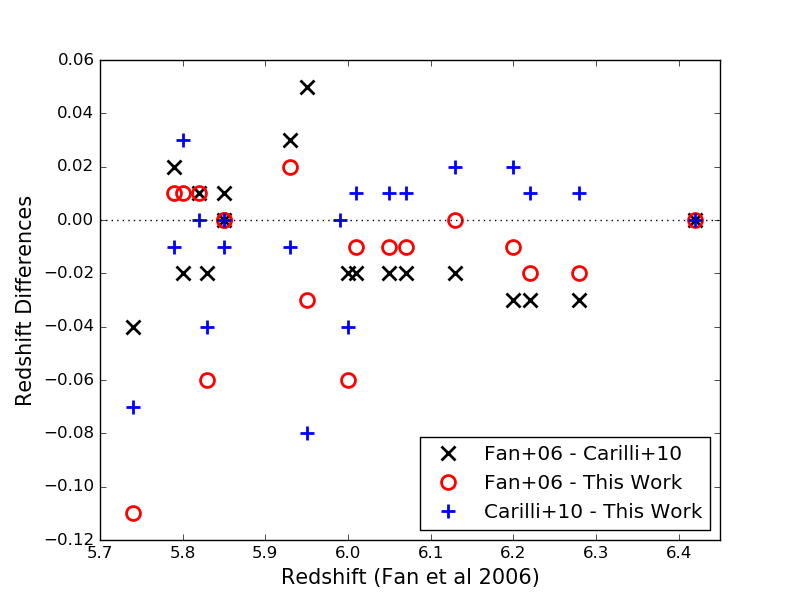}
\caption{A comparison of the differences in redshifts between the fitting
method used
here and the results from \citet{Fan2006b} and \citet{Carilli2010}.
The dashed line indicates the zero line.} \label{zComp} \end{figure}

The method was tested on the SDSS sample from \citet{Fan2006b}; there it was
found to recover the redshifts presented with a median difference of -0.01
with $\sigma_{\mathrm{MAD}} = 0.01$. The $\sigma_{\mathrm{MAD}}$ (median
absolute deviation (MAD)) is used as a robust estimator of the gaussian
standard deviation where $\sigma_{\mathrm{MAD}} = 1.4826 \times \mathrm{MAD}$.
$\sigma_{\mathrm{MAD}}$ was used to give an estimate of the systematic
uncertainty in the redshifts of 0.01 which is far larger the statistical
uncertainties from the fitting.  As the data quality varies across the sample
the uncertainties are going to be underestimated for the nosiest data. The
calculated redshifts and the redshifts from \citet{Fan2006b} were also compared
with the redshifts presented in \citet{Carilli2010}; as shown in Figure
\ref{zComp}. The median difference between our calculated redshifts and the
redshifts from \citet{Carilli2010} was found to be 0.0 with
$\sigma_{\mathrm{MAD}} = 0.01$ while the median difference between the
redshifts from \citet{Carilli2010} and \citet{Fan2006b} was -0.02 with
$\sigma_{\mathrm{MAD}} = 0.01$.

\begin{landscape}
\begin{table}
\begin{center}
\caption{Properties of the quasars in this sample.
Upper limits are given for the magnitude in a 2" aperture. All magnitudes are
given in AB.}
\label{tab:properties}
\begin{tabular}{lccccccccccccc}
\hline
Name& Ranking & DES      & RA      & DEC     &  g & r & i & z & Y & J & Ks & W1 & W2 \\
    &         & Tilename & (J2000) & (J2000) &    &   &   &   &   &   &    &    &    \\
\hline
VDESJ0143-5545 & 9 & DES0145-5540 & 25.79265 & -55.75297 & $>$ 24.22 & $>$ 23.83 & 24.03 & 20.50 & 21.11 & 20.61 & 20.09 & 19.39 & 19.00 \\
& & & 01$^{h}$43$^{m}$10.24$^{s}$ & -55$^{\circ}$45'10.68"&         &           & $\pm$ 0.19 & $\pm$ 0.02 & $\pm$ 0.12 & $\pm$ 0.11 & $\pm$ 0.18 & $\pm$ 0.09 & $\pm$ 0.10 \\[5pt]

VDESJ0224-4711 & 3 & DES0222-4706 & 36.11057 & -47.19149 & $>$ 23.47 & $>$ 23.35 & 24.02      & 20.20      & 19.89      & 19.75      & 18.99      & 18.75 & 18.64 \\
& & & 02$^{h}$24$^{m}$26.54$^{s}$ & -47$^{\circ}$11'29.4"&          &           & $\pm$ 0.40 & $\pm$ 0.02 & $\pm$ 0.05 & $\pm$ 0.06 & $\pm$ 0.06 & $\pm$ 0.05 & $\pm$ 0.14 \\[5pt]

VDESJ0323-4701 & 10 & DES0325-4706 & 50.91808 & -47.02226 & $>$ 24.40 & $>$ 24.10 & 24.30      & 20.78      & 20.74      & 20.88      & 20.51      & 20.31      & 20.51 \\
& & & 03$^{h}$23$^{m}$40.34$^{s}$ & -47$^{\circ}$01'20.13"&         &           & $\pm$ 0.26 & $\pm$ 0.02 & $\pm$ 0.07 & $\pm$ 0.16 & $\pm$ 0.26 & $\pm$ 0.17 & $\pm$ 0.30 \\[5pt]

VDESJ0330-4025 & 5 &  DES0329-4040 & 52.61632 & -40.42121 & 24.98      & $>$ 23.80 & 23.76      & 20.66      & 20.34      & 20.56      & 19.99      & 19.55      & 19.58 \\
& & & 03$^{h}$30$^{m}$27.92$^{s}$ & -40$^{\circ}$25'16.4"& $\pm$ 0.37 &           & $\pm$ 0.20 & $\pm$ 0.02 & $\pm$ 0.06 & $\pm$ 0.13 & $\pm$ 0.18 & $\pm$ 0.09 & $\pm$ 0.14 \\[5pt]

VDESJ0408-5632 & 8 & DES0407-5622 & 62.08012 & -56.54134 & $>$ 24.04 & 24.89 & 22.48 & 20.13 & 20.19 & 19.91 & 19.70 & 20.30 & 19.74 \\
& & & 04$^{h}$08$^{m}$19.23$^{s}$ & -56$^{\circ}$32'28.82"&          & $\pm$ 0.42 & $\pm$ 0.10 & $\pm$ 0.01 & $\pm$ 0.05 & $\pm$ 0.06 & $\pm$ 0.14 & $\pm$ 0.15 & $\pm$ 0.13 \\

VDESJ0410-4414 & 1 & DES0409-4414 & 62.51345 & -44.24464 & $>$ 24.01 & 25.31   & $>$ 23.04 & 20.65      & 20.61      & 20.68      & 20.22      & 20.00      & 20.79 \\ 
& & & 04$^{h}$10$^{m}$03.23$^{s}$ & -44$^{\circ}$14'40.7"&       & $\pm$ 0.45 &           & $\pm$ 0.02 & $\pm$ 0.09 & $\pm$ 0.13 & $\pm$ 0.22 & $\pm$ 0.12 & $\pm$ 0.34 \\[5pt]

VDESJ0420-4453 & 6 & DES0421-4457 & 65.04727 & -44.88993 & $>$ 24.27 & 24.98      & 22.71      & 20.32      & 20.46      & 20.57      & 20.04      & 19.73      & 20.12 \\
& & & 04$^{h}$20$^{m}$11.34$^{s}$ & -44$^{\circ}$53'23.8"&          & $\pm$ 0.27 & $\pm$ 0.07 & $\pm$ 0.02 & $\pm$ 0.06 & $\pm$ 0.12 & $\pm$ 0.19 & $\pm$ 0.10 & $\pm$ 0.21 \\[5pt]

VDESJ0454-4448$^{\dagger}$ & 2 & DES0453-4457 & 73.50744 & -44.80864 & $>$ 24.46 & $>$ 24.09 & 22.64      & 20.24      & 20.36      & 20.24      & 20.11      & 19.62      & 19.70 \\
& & & 04$^{h}$54$^{m}$01.79$^{s}$ & -44$^{\circ}$48'31.1"&       &           & $\pm$ 0.05 & $\pm$ 0.01 & $\pm$ 0.05 & $\pm$ 0.07 & $\pm$ 0.18 & $\pm$ 0.10 & $\pm$ 0.15 \\[5pt]

VDESJ2250-5015 & 4 & DES2250-4957 & 342.50837 & -50.26171 & $>$ 23.60 & $>$ 23.68 & 22.63 & 20.11 & 19.98 & 19.18 & 19.00 & 18.71 & 19.04 \\
& & & 22$^{h}$50$^{m}$02.01$^{s}$ & -50$^{\circ}$15'42.15"&         &           & $\pm$ 0.06 & $\pm$ 0.01 & $\pm$ 0.04 & $\pm$ 0.21 & $\pm$ 0.14 & $\pm$ 0.06 & $\pm$ 0.12 \\[5pt]

VDESJ2315-0023$^\star$ & 7 & DES2316-0041 & 348.94409 & -0.39938 & 24.00 & $>$ 23.62 & 23.81 & 20.83 & 21.54 & 21.47 & - & 20.65 & - \\
& & & 23$^{h}$15$^{m}$46.58$^{s}$ & 00$^{\circ}$23'57.78"& $\pm$ 1.86 & & $\pm$ 0.20 & $\pm$ 0.03 & $\pm$ 0.30 & $\pm$ 0.22 & & $\pm$ 0.31 &  \\[5pt]

\hline
\vspace{-8pt}\\
\multicolumn{11}{l}{\scriptsize{$^{\dagger}$ This quasar was found in R15 and it is included here for
completeness and to allow comparison between the different DES data releases
and the different \textit{WISE} reductions used.}}\\
\multicolumn{11}{l}{\scriptsize{$^\star$ This is a known object
(SDSS J231546.57-002358.1) discovered in \citet{Jiang2008}}}\\

\end{tabular}
\end{center}
\end{table}
\end{landscape}

\begin{table*}
\begin{center}
\caption{Derived properties of the quasars in this sample.
The near zone sizes for VDESJ0454-4448 are taken
from R15. Near zone sizes are not given for all objects as the data quality
was not good enough.} \label{tab:DerivedProperties} \begin{tabular}{lcccc}
\hline
Name & Redshift & M$_{\mathrm{1450}}$ & R$_{NZ}$ & R$_{NZ, corrected}$\\
\hline
VDESJ0143-5545 & 6.25 $\pm$ 0.01 & -25.65 $\pm$ 0.12 & & \\
VDESJ0224-4711 & 6.50 $\pm$ 0.01 & -26.93 $\pm$ 0.05 & & \\
VDESJ0323-4701 & 6.25 $\pm$ 0.01 & -26.02 $\pm$ 0.07 & 2.1 $_{-0.5}^{+0.6}$ Mpc & 2.8 $_{-0.7}^{+0.8}$ Mpc \\
VDESJ0330-4025 & 6.25 $\pm$ 0.01 & -26.42 $\pm$ 0.06 & 2.1 $_{-0.5}^{+0.6}$ Mpc & 2.5 $_{-0.6}^{+0.7}$ Mpc \\
VDESJ0408-5632 & 6.03 $\pm$ 0.01 & -26.51 $\pm$ 0.05 & & \\
VDESJ0410-4414 & 6.21 $\pm$ 0.01 & -26.14 $\pm$ 0.09 & 6.9 $_{-0.5}^{+0.5}$ Mpc & 9.0 $_{-0.7}^{+0.6}$ Mpc \\ 
VDESJ0420-4453 & 6.07 $\pm$ 0.01 & -26.25 $\pm$ 0.06 & 4.3 $_{-1.0}^{+0.6}$ Mpc & 5.3 $_{-1.2}^{+0.8}$ Mpc \\
VDESJ0454-4448$^{\dagger}$ & 6.10 $\pm$ 0.01 & -26.36 $\pm$ 0.05 & 4.1 $_{-1.2}^{+1.1}$ Mpc & 4.8$_{-1.4}^{+1.3}$ Mpc\\ 
VDESJ2250-5015 & 6.00 $\pm$ 0.01 & -26.80 $\pm$ 0.04 & & \\
\hline
\vspace{-8pt}\\
\multicolumn{4}{l}{\scriptsize{$^{\dagger}$This object was found in
R15.}}\\ 
\end{tabular}
\end{center}
\end{table*}

%\section{Results}

\section{Quasar ionization near zones}

\begin{figure} 
\includegraphics[width = \linewidth]{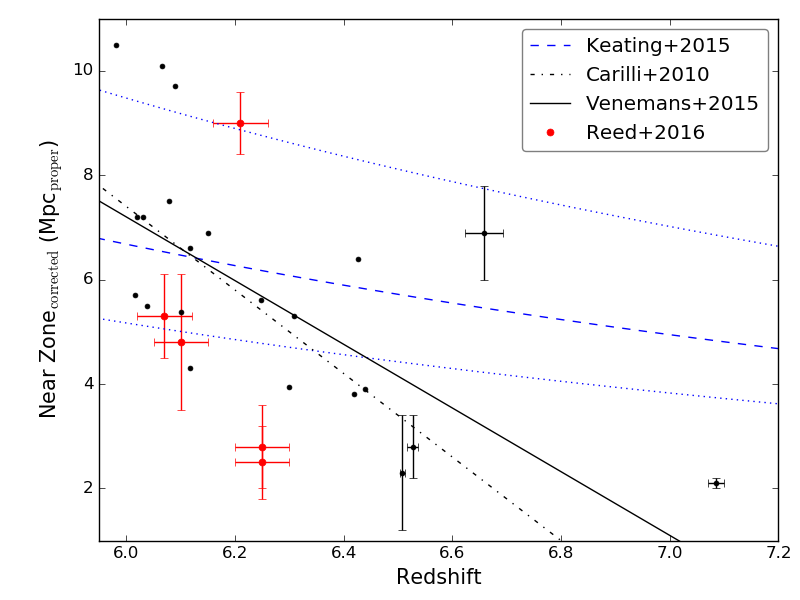}
\caption{A comparision of the theoretical predictions and observations for high
redshift quasar near zone sizes. The black line shows the fit to the
observational data from \citet{Carilli2010} and the black dot-dashed line is
the fit from \citet{Venemans2015}. The blue line shows the theoretical fit from
\citet{Keating2015} and the blue dotted lines are the 25$^{\textrm{th}}$ and
75$^{\textrm{th}}$ percentile for the range of near zones sizes that they
found. The black points show near zone sizes from known quasars in the
literature. The red points are some of the quasars in this sample. Objects with
poor signal to noise spectra were not included here.}
\label{Figure:NearZones} \end{figure}

The observed spectra of z $>$ 6 quasars are characterised by intrinsic quasar
continuum emission and emission lines longward of the Lyman-$\alpha$ emission
line in the quasar rest frame. Shortward of Lyman-$\alpha$ in the quasar rest
frame the spectrum the most distinctive feature is the deficit of continuum
emission due to HI Lyman-$\alpha$ and Lyman series absorption absorption by the
cosmologically distributed intervening Lyman-$\alpha$ forest. At z $>$ 6 the
optical depth from this neutral HI absorption is considerable and is often
called the 'Gunn Peterson trough' where the neutral hydrogen fraction
($f_{\mathrm{HI}}$) is $f_{\mathrm{HI}} > 10^{-3}$. Closer to the quasar the UV
radiation from the quasar ionizes HI and and the HI opacity is decreased. This
highly ionized HII region in called a near zone and the size of this region is
determined by the large scale structure or clumpiness of the HI, the average
neutral fraction, the UV luminosity of the quasar and the age of the expanding
UV radiation front emitted by the quasar. Observations of the distibution of
near zone sizes and the evolution with redshift of this distribution is an
important probe of the Universe in the epoch of reionization. 

Near zones sizes were calculated using the method described in R15 which 
follows \citet{Fan2006b} where the edge of the near zone is taken
to be the point where the ratio between the continuum flux and the spectra
first falls below 0.1 blueward of the Ly$\alpha$ peak. The spectral resolution
and signal to noise of our four NTT spectra are too low to measure near zones
sizes. Measured near zone sizes ($\rm R_{NZ}$) measurements from the four 
Gemini spectra and from R15 are presented in Table \ref{tab:DerivedProperties}.
The near zone size of a quasar in an cosmologically expanding medium
will depend on the intrinsic UV flux of the quasar below the Lyman-$\alpha$ 
transtion at 1216$\mathrm{\AA}$. Following \citet{Carilli2010} we normalise the
measured near zone sizes  ($\rm R_{NZ}$) to a constant UV absolute magnitude 
$M_{1450} = -27$ with the equation below.

$$R_{\mathrm{NZ, corrected}} = R_{\mathrm{NZ}} \times 10^{0.4(27.0+M_{1450})/3}$$

Figure \ref{Figure:NearZones} shows the distribution of corrected near zone
size for 18 quasars with $6.0<z<6.5$ from \citet{Carilli2010}  and 4 z$>$6.5
quasars from \citet{Venemans2015} and \citet{Mortlock2011} along with our new
sample with $6.0<z<6.5$. 

The blue solid line is the analytic solution from \cite{Keating2015} for the
evolution of the normalised near zone sizes with redshift where the quasar has
constant luminosity and the neutral fraction is not evolving with redshift. The
decrease in size with increasing redshift is solely due to the increase in mean
HI density as the Universe gets smaller in size at earlier  redshifts. The
dashed blue lines show the 15th and 85th percentiles about the median ($\sim
\pm 1 \sigma$) derived from simulations (\cite{Keating2015}).  The black dashed
and black dot dashed lines show linear fits by \citet{Carilli2010}  and
\citet{Venemans2015} respectively. 

The four new zones that we measure at $6.1 < z < 6.3$ span a large range from 3
to 9 Mpc. Two, VDES J0330-4025 and VDES J0323-4701, have relatively
small corrected near zone sizes of $\sim$3 Mpc which could indicate that these
two quasars are younger than the average quasar at this epoch and have
relatively small lifetime ($10^6 - 10^7$ years) and the ionized HII regions have
not reached their maximum size due to the time taken for the ionizing radiation
fronts to expand into the surrounding HI region.  Alternatively if one ignores
the effects of quasar lifetime to fully account for the small near zone sizes
the objects would need to be situated in regions of the Universe which are a
factor of $\sim$10 above average HI density.  Similar effects has been reported
by \citet{Bolton2011} for the z = 7.085 quasar ULAS J1120+0641. The discovery of
two z$\sim$6.2 quasars with such small near zones indicates that care needs to
taken in interpreting small near zones as evidence for an increase in the
neutral fraction. To further address this more observational data is essential.

\section{Properties of Individual Objects}

\begin{figure}
\includegraphics[width = 1.0\linewidth]{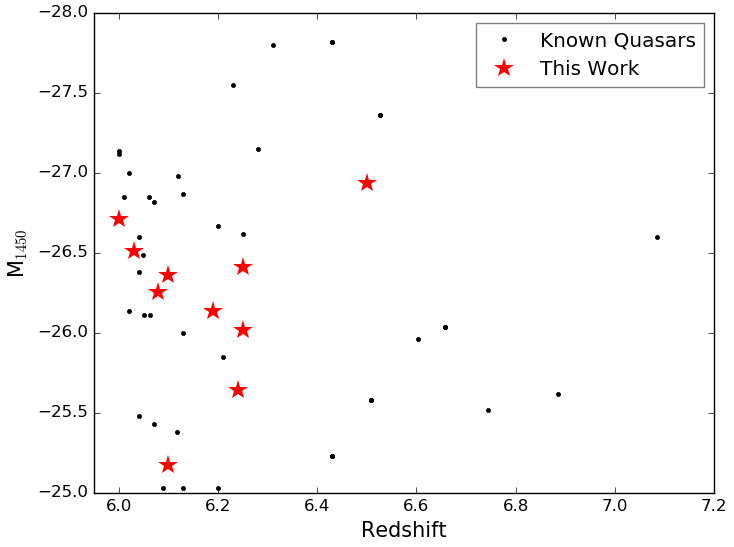}
\caption{Here the absolute magnitude calculated at 1450$\mathrm{\AA}$ in the
restframe is shown against redshift. The M$_{1450}$ was estimated from the
Y band magnitude of the objects.}
\label{M1450_z}
\end{figure}

\begin{figure}
\includegraphics[width = 1.0\linewidth]{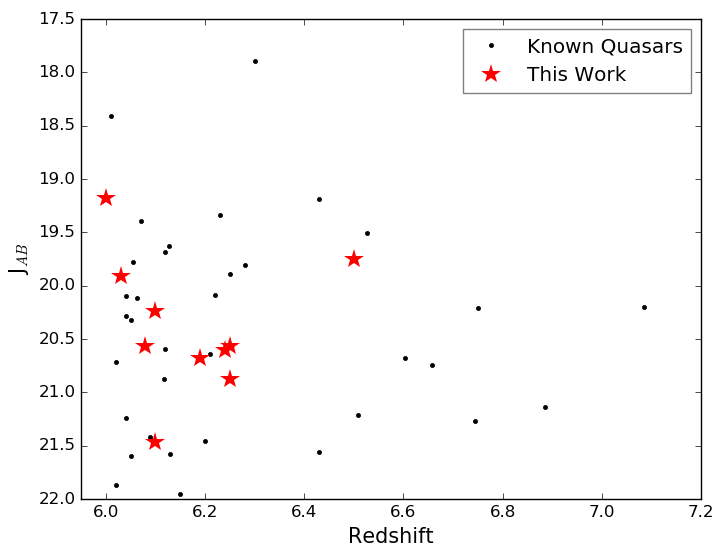}
\caption{The apparent AB magnitude of these quasars in the J band is shown
against their redshifts compared with known quasars.}
\label{JAB_z}
\end{figure}

Here we give more details on some specific objects from our sample. A summary
of the derived properties of the quasars presented here is given in Table
\ref{tab:DerivedProperties}.  A comparison of these to known quasars is shown
in Figures \ref{M1450_z} and \ref{JAB_z}.

\label{IndObjects}
%\subsection{VDESJ0410-4414}
%The candidate ranked 1$^{st}$ by the code was followed up with GMOS and the
%distinctive break in the spectra was seen at $\sim$8720$\mathrm{\AA}$ giving
%a quasar at z $\sim$ 6.2 matching the predicted redshift of 6.2.  This object
%was completely undetected in the i band and has blue colours in z - Y,
%Y - J (z-Y = 0.05 $\pm$ 0.09, Y - J = -0.08 $\pm$ 0.16).

\subsection{VDESJ0143-5545 (z = 6.23)}
J0143$-$5545 was followed up with the NTT and found to have a very strong
emission feature at $\sim$8820$\mathrm{\AA}$ suggestive of a quasar with z
$\sim$ 6.3. 
This object was well fit by the model with the highest level of
reddening, E(B-V) = 0.100, at z = 6.1. 
This object has a very blue z - Y of -0.61
due to the presence of the very strong Ly$\alpha$ emission line in the z
filter. When the reddening fit was repeated without using the blended
\textit{WISE} data the object was best fit by a model with E(B-V) = 0.025
suggesting that the W1 and W2 fluxes are effected by a nearby source.

\subsection{VDESJ0224-4711 (z = 6.50)}
This candidate was ranked as the third most likely object to be a quasar in the
candidate list with a very good fit to a reddened quasar model (E(B-V) = 0.05)
at z $\sim$ 6.4.  It is quite bright with z = 20.0 and has a very red i$-$
z colour of 3.82. Follow up of this object with the NTT showed a strong
emission feature starting at $\sim$9100$\mathrm{\AA}$ giving a redshift of
6.50.  The reddening fit was recalculated with the redshift fixed at the
observed spectroscopic redshift of 6.50.  At the spectroscopic redshift the
photometry was best fit by a reddened model with E(B-V) = 0.05.  This object
appears to have a very extended near zone but the modest quality of the
spectral data means that this measurement has very large uncertainties.
VDESJ0224$-$4711 has J$_{AB}$ = 19.75 is the second most luminous quasar known
with z$\geq$6.5 and is 0.2mag fainter than the most luminous quasar know with
z$>$6.5; PSO J0226+0302 with z = 6.53 and J$_{AB}$ = 19.51
\citet{Venemans2015}.

\subsection{VDESJ0323-4701 (z = 6.25)}
VDESJ0323$-$4701 was the lowest ranked candidate
followed up.  Spectroscopic observations with GMOS revealed a quasar at
z $\sim$ 6.25. This object was very red with i - z = 3.52 and was best fit by
a non reddened quasar model with a slightly lower redshift of 6.10 than the
spectroscopic one. 
The measured corrected near zone size of 2.8 proper Mpc 
which could indicate above averge IGM density of a young age for this
quasar.

\subsection{VDESJ0330-4025 (z = 6.25)}
The measured corrected near zone size of 2.5 proper Mpc 
which could indicate above averge IGM density of a young age for this
quasar. VDESJ0330-4025 lies within 10 deg on the sky of the other
quasar with a small near zone; VDESJ0323-4701 which also has a redshift
of 2.5 so they could lie in a correlated region of the Universe with 
above average IGM over density.

\subsection{VDESJ0454$-$4448 (z = 6.10)}
VDESJ0454$-$4448 was the first object identified from DES and was the subject of
R15; details of the spectroscopic observations are included therein. It is
included here as it was covered again by the year one release from the DES and
was the second highest ranked candidate in the independent 
data analysis in this paper. The redshift was
recalculated for this object as part of this analysis and was found to be 6.10
$\pm$ 0.01 which is consistent with the value given in R15 of 6.09 $\pm$
0.03.

\subsection{VDESJ2250$-$5015 (z = 6.00}
This object was ranked fourth by the selection code with a good fit to
a model with E(B$-$V) = 0.05 and a predicted redshift of 6.0. Follow up
spectroscopy with the NTT gives a redshift of 6.00. This source was the
brightest in our sample with z = 20.11. VDESJ2250-5015 has a fairly red i - z
colour of 2.52 and has a very red Y - J colour of 0.80. The reddening fit was
repeated with the redshift fixed at the calculated one and without using the
\textit{WISE} data as the close proximity of another source might be
influencing this. This resulted in a model with more reddening (E(B-V) = 0.1)
being chosen as the best fit. The red Y - J colour of this object is probably
due to the reddening.

%\subsection{VDESJ0330-4025}
%VDESJ0330-4025 was ranked as the fifth most likely candidate. This object was
%also very red in i $-$ z (i $-$ z = 3.10) giving a reasonably high predicted
%redshift of 6.2. A spectrum of this object was taken with GMOS and the strong
%feature at $\sim$8820$\mathrm{\AA}$ is indicative of a quasar at z = 6.25.
%There is also a tentative detection of Ly-$\beta$ at $\sim$7450$\mathrm{\AA}$.

%\subsection{VDESJ0420-4453}
%The sixth object was followed up with GMOS and the spectra showed emission
%identified as Ly-$\alpha$ at $\sim$8600$\mathrm{\AA}$ giving z $\sim$ 6.1. The
%spectra shows potential Ly-$\beta$ emission at $\sim$7270$\mathrm{\AA}$. This
%source also had blue colours in z-Y, Y-J and W1-W2.

\subsection{VDESJ2315-0023 (z = 6.12)}
The seventh mostly likely ranked candidate was a known quasar 
(SDSS J231546.57-002358.1) from the SDSS survey for quasars in stripe 82
\citep{Jiang2008}. Their spectropscopic follow up found it to have z = 6.117
which is slightly higher than our photometric estimate of 6.0.

%\subsection{VDESJ0408-5632}
%The object ranked eigth was followed up with the NTT was VDESJ0408-5632. The
%spectrum shows a sharp break at $\sim$8540$\mathrm{\AA}$ suggestive of a quasar
%at 6.0 which agrees with the best fitting photometirc redshift. This source is
%the second brightest of our sample at z = 20.13.

\section{Analyis of Selection Method}

There are seven prevoiusly known quasars with z $\geqslant$ 5.80 in the area
covered by the data used in this study. Two are recovered by the selection
criterion, VDES0454-4448 (z = 6.10) from R15 and SDSSJ2315+0023 from Jiang et
al, 2008 as discussed in section \ref{IndObjects}.  SDSS J000552.34000655.8 (z
= 5.85) discovered in \citet{Fan2004} is bluer than our i$_{\mathrm{DES}}$
- z$_{\mathrm{DES}}$ selection and is not selected.  This colour
is indicative of being at a lower redshift than this selection method probes.
The three quasars in \citet{Jiang2009} and the radio selected z = 5.95 quasar
(SDSS J222843.54+011032.2) \citet{Zeimann2011} that overlap the area have
z$_{\mathrm{DES}}$ $>$ 21.0 and therefore are fainter than our selection limit.

The automatic ranking of candidates in the candidate list allows visual
inspection to be prioritised.  This will be particularly useful once the full
DES area is available for study as there will be a large number ($\sim$500) of
candidate objects.  This also means that looser colour cuts can be used to
narrow down the data slightly allowing more unusual objects to be discovered.
One such object is VDESJ2250-5015 whose red colour in Y-J would have caused
it to be rejected by previous searches (\citealt{Banados2016, Venemans2015b}).
The Y and J band photometry of VDESJ2250-5015 is reliable and suggests that the
very red colour is real and due to intrinsic properties of the object.

The SED model fitting selection method presented here also allows the expansion
of the candidate list without increasing the need for visual inspection as
objects can be double checked in the ranked order. This is because most of the
types of junk that contaminate the list are classified as very unlikely to be
quasars.  Our future aim is to be able to run the selection criteria over the
entire input list without any need for colour cuts and using the reduced
$\chi^{2}$ fits as discriminators. At the moment the colour selection is
required to narrow down the list enough to make the image based steps run more
rapidly. Improvements in the analysis code will allow this to be done for
a larger number of images more rapidly. The catalogue based steps and the
fitting steps are both fast enough ($10^8$ sources from $\sim$1500 deg$^2$
sources in less than 24hr on a single 4Ghz core) that they will be easily
expanded to the larger $\sim$5000 deg$^{2}$ DES dataset when it is released.

\begin{figure} 
\includegraphics[width = \linewidth]{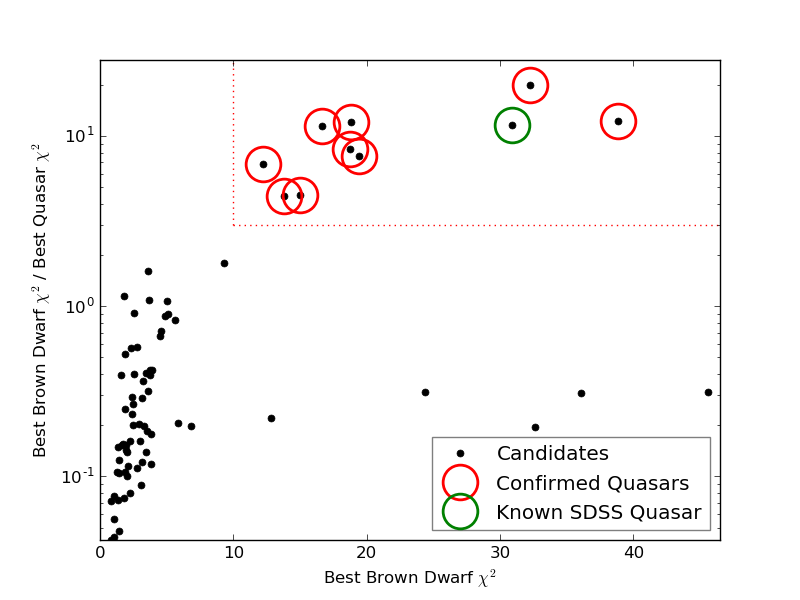}
\caption{The known quasars were used to derive cuts to automatically select
quasar candidates. These cuts are shown as the red dotted lines. The points
circled in black are DES quasars and the object circled in green is a known
SDSS quasar. It can be seen that these cuts separate the known quasars well
from the rest of the sample. Due to inclement weather the object between the
locus of points and the delimited box has not been followed up.}
\label{Figure:Chi2BD} \end{figure}

In this version of the selection code objects with a high probability of being
a brown dwarf are not rejected automatically but removed on an object by object
basis when image cutouts of the object are checked.  An improvement to the
method would be to have automatic removal of these objects, as this will be
more important for larger candidate lists generated either by relaxing the
colour cuts or by a larger input dataset. The confirmed quasars compared to the
rest of the sample are shown in terms of reduced $\chi^{2}$ to be either a star
or a quasar in Figure \ref{Figure:Chi2BD}. It can be seen that the selected
objects are well separated from the rest of the sample. Due to the inclement
weather we did not have time to follow up any objects further down the ranking
so do not know if the dashed lines should be relaxed to select a complete
sample. The candidates in the bottom right region are junk as confirmed by
visual inspection.
%This is discussed further in section \ref{NoCC}.

\begin{figure} \includegraphics[width = \linewidth]{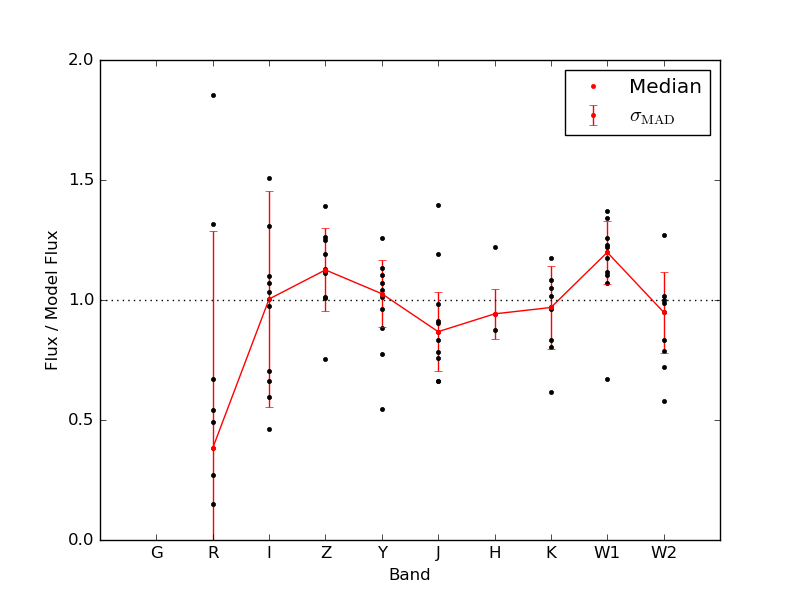}
\caption{The black points are the measured flux values for the sample 
of 10 quasars
divided by the flux derived from the best fitting model. The red points 
are the medians of the ratio values in each band. The red 
error bars show the 1$\sigma$ uncertainty derived
from the median absolute deviation. No g band is shown as the model flux was
mostly zero.} \label{Figure:Residuals}
\end{figure}

In Section \ref{Section:PhotoZ} our implemented SED fitting included an
arbitrary systematic flux uncertainty. Now we have a spectroscopically
confirmed sample we analyze the residuals from the best fit models. In Figure
\ref{Figure:Residuals} we show the ratio of the observed flux to the best fitting
model flux for each quasar, the sample median and the $\rm\sigma_{MAD}$. These
show that the best fit models agree within the uncertainties.

The scatter in the ratios as described by the $\rm\sigma_{MAD}$ is shown in
Figure \ref{Figure:Residuals}. The values for these ratios in the bands that
are unaffected by the Ly$\alpha$ forest are similar to the values assumed in
the fitting as described in Section \ref{Section:PhotoZ}. In r and i the large
scatter is from stocastic scatter in Lyman-$\alpha$ forest and phometric
statistical errors. In a future paper with a larger sample of confirmed quasars
we will investigate the scatter in terms of the model.  There is some evidence
for excess flux in the W1 band which is probably due to the large aperture used
resulting in flux from neighbouring objects. Since quasars are redder in
W1 - W2 than the foreground galaxy and stellar populations, the W2 band is less
effected.

% Percentage errors in flux of 10%, 10%, 10%, 20%, 5%, 5%, 312
% 292 5%,5%,20%and20%ing,r,i,z,Y,J,H,Ks,W1andW2respec- 313
% 293 tively were added in quadrature to the statistical uncertainties.

The code is written to allow different models to be easily inserted and tested
meaning that as additional models become available we can also compare to
these. This will allow us to search for a wider variety of quasar types. For
example we hope to include models with a wider range of extinctions and at
a finer sampling in extinction and redshift. Different treatments of the 
Ly$\alpha$ forest and different properties of the IGM can also be incorporated.

\section{Summary and Conslusions}

%We have presented a new method of selecting high redshift quasars that
%uses all the available photometry and use SED fitting of 
%a range of quasar and brown 
%dwarf models to all the
%available photometric data for each candidate object has been developed. This
%allows prioritisation of candidates for follow up and offers a very high
%success rate for confirming candidates spectroscopically allowing telescope
%time to be used efficiently. This method also allows objects with more unusual
%properties to be discovered due to the range of reddening incorporated into the
%models. This method is fast enough to be used on large datasets after loose
%colour cuts have been applied and will be easily applied to the next release of
%DES data. In the future  the metid can also be extended to higher redshifts 
%as the models extend to z$\sim$8 and deeper optical and near IR data comes
%available LSST and Euclid.

We have presented the photometric selection, statistical classification and
spectroscopic confirmation of eight new high redshift 
6.0 $<$ z $<$ 6.5 quasars with z$_{\mathrm{AB}}$ $<$ 21.0, selected without any
morphological star-galaxy classification from $\sim$1500 deg$^{2}$ using SED
model fitting to photometric data from the Dark Energy Survey (g, r, i, z, Y),
the VISTA Hemisphere Survey (J, H, K) and the Wide-Field Infrared Survey
Explorer (W1, W2). Starting from over 100 million photometric sources we used
objective and repeatable machine based techniques to select 147 quasar
candidates.  Our spectral confirmation success rate is 100\% without the need
for follow-up photometric observations as used in other studies of this type.
Combined with automatic removal of the main types of non-astrophysical
contaminants the method allows large data sets to be processed without human
intervention and without being over run by spurious false candidates.  The
highest redshift quasars with z = 6.5; VDESJ0224$-$4711 which has J$_{\mathrm{AB}}$
= 19.75 is the second most luminous quasar known with z $\geq$ 6.5 and is 0.2 mag
fainter than the most luminous quasar know with z$>$6.5; PSO J0226+0302 with
z = 6.53 and J$_{AB}$ = 19.51 \citet{Venemans2015}.

Candidates were ranked based on the ratio of reduced $\chi^2$-statistic values
for the best fit quasar model compared to the best fit stellar model.
This is is approach is extendable to other photometric systems and imaging surveys 
(e.g. LSST, Euclid), in contrast to colour cut based criteria widely
used in other high redshift quasars searches.

A new quasar redshift determination algorithm has been developed based on the 
onset of the Lyman-$\alpha$ forest and a fit to the Lyman-$\alpha$ emssion
line using a semi-Gaussian and an exponential. The technique is validated  
on a sample of quasar that also have CO and MgII emission line redshifts from
\citet{Carilli2010} and find that our empirical fitting technique has a median
difference of 0.003 and the distribution has $\sigma_{\mathrm{MAD}}$ = 0.01. 

We have measured the sizes of the quasar ionization near zones for 4 of
the new quasars and the z = 6.00 quasar J0454-4448 from \citep{Reed2015} as
shown Figure \label{Figure:NearZones}.  The four new zones that we measure at
$6.1 < z < 6.3$ span a large range from 3 to 9 Mpc. Two, VDES J0330-4025
and VDES J0323-4701, have relatively small corrected near zone sizes of
$\sim$3 Mpc which could indicate that these two quasars are younger than the
average quasar at this epoch and have relatively small lifetime ($10^6 - 10^7$
years) and the ionized HII regions have not reached their maximum zone size due
to the time taken for the ionizing radiation fronts to expand into the
surrounding HI region.  Alternatively if one ignores the effects of quasar
lifetime to fully account for the small near zone sizes the objects would need
to be situated in regions of the Universe which are a factor of $\sim$10 above
average HI density.  Similar effects has been reported by \citet{Bolton2011}
for the z = 7.085 quasar ULAS J1120+0641. The discovery of two z$\sim$6.2 quasars
with such small near zones indicates that care needs to taken in interpreting
small near zones as evidence for an increase in the neutral fraction. To
further address this more observational data is essential.

%These quasars were photometrically selected without the use of
%any star-galaxy morphological
%criteria from 1523
%deg$^{2}$ using SED model fitting to photometric data from the Dark
%Energy Survey (g, r, i, z, Y), the VISTA Hemisphere Survey (J, H,
%K) and the Wide-Field Infrared Survey Explorer (W1, W2). The photometric data
%was fitted with a grid of quasar model spectral energy 
%distributions with redshift dependent
%Lyman-$\alpha$ forest absorption and a range of levels of intrinsic reddening
%as well as a series of low mass cool star stellar models. 
%Candidates were ranked based
%on a SED-modelling based $\chi^2$-statistic, 
%which is extendable to other photometric systems and imaging surveys 
%(e.g. LSST, Euclid),
%rather than conventional 
%colour cut criteria. 
%which
%allows more unusual objects to be found such as VDESJ2250$-$5015 which has 
%red
%colours (E(B-V)=0.1) and would have be missed by previous surveys. 

We also present a robust parametric redshift estimating technique based on the
onset of the Lyman-alpha forest that gives compararable accuracy MgII and CO
based redshift estimators.

\section{Acknowledgements}
SLR and RGM would like to thank Laura Keating for an interesting and
educational discussion on near zones.  
RGM, SLR, MB, MA, PH, SEK, SLJG and EGS
acknowledge the support of UK Science
and Technology Facilities Council (STFC). Support for RGM by 
ERC Advanced Grant 320596 'The Emergence of Structure During the Epoch of 
Reionization' is gratefully acknowledged.
MB acknowledges funding from STFC via an Ernest Rutherford Fellowship.

Funding for the DES Projects has been provided by the US Department of Energy,
the US National Science Foundation, the Ministry of Science and Education of
Spain, the Science and Technology Facilities Council of UK, the Higher
Education Funding Council for England, the National Center for Supercomputing
Applications at the University of Illinois at Urbana-Champaign, the Kavli
Institute of Cosmological Physics at the University of Chicago, Financiadora
de Estudos e Projetos, Funda{\c c}{\~a}o Carlos Chagas Filho de Amparo {\'a}
Pesquisa do Estado do Rio de Janeiro, Conselho Nacional de Desenvolvimento
Cient{\'i}fico e Tecnologico and the Minist{\'e}rio da Ci{\^e}ncia
e Tecnologia, the Deutsche Forschungsgemeinschaft and ˆ the Collaborating
Institutions in the Dark Energy Survey.

The Collaborating Institutions are Argonne National Laboratories, the
University of California at Santa Cruz, the University of Cambridge, Centro de
Investigaciones Energeticas, Medioambientales y Tecnologicas-Madrid, the
University of Chicago, University
College London, the DES-Brazil Consortium, the Eidgenossische Technische
Hochschule (ETH) Zurich, Fermi National Accelerator Laboratory, the University
of Edinburgh, the University of Illinois at Urbana-Champaign, the Institut de
Ciencies de l’Espai (IEEC/CSIC), the Institut de Fisica d’Altes Energies, the
Lawrence Berkeley National Laboratory, the Ludwig-Maximilians Universit{\"a}t
and the associated Excellence Cluster Universe, the University of Michigan, the
National Optical Astronomy Observatory, the University of Nottingham, The Ohio
State University, the University of Pennsylvania, the University of Portsmouth,
SLAC National Laboratory, Stanford University, the University of Sussex, and
Texas A\&M University.

The analysis presented here is based on observations obtained as part of the
VISTA Hemisphere Survey, ESO Progamme, 179.A-2010 (PI: McMahon).
The analysis presented here is based on observations obtained as part of ESO
Progamme, 096.A-0411 (PI: McMahon) and GEMINI programme GS-2015B-Q-14 (PI:
Martini).

This analysis makes use of the cosmics.py algorithum based off Pieter van
Dokkum's L.A. Cosmic algorithum detailed in \citet{VanDokkum2001}.

This paper has gone through internal review by the DES collaboration.

%ACR acknowledges financial support provided by the PAPDRJ CAPES/FAPERJ
%Fellowship

\bibliographystyle{mn2e}
\bibliography{Refs}

\section*{Affiliations}
{\small
$^{1}$Institute of Astronomy, University of Cambridge, Madingley Road, Cambridge CB3 0HA, UK\\
$^{2}$Kavli Institute for Cosmology, University of Cambridge, Madingley Road, Cambridge CB3 0HA, UK\\
$^{3}$Department of Physics \& Astronomy, University College London, Gower Street, London, WC1E 6BT, UK\\
$^{4}$Center for Cosmology and Astro-Particle Physics, The Ohio State University, Columbus, OH 43210, USA\\
$^{5}$Department of Astronomy, The Ohio State University, Columbus, OH 43210, USA\\
$^{6}$Department of Physics and Electronics, Rhodes University, PO Box 94, Grahamstown, 6140, South Africa\\
$^{7}$Fermi National Accelerator Laboratory, P. O. Box 500, Batavia, IL 60510, USA\\
$^{8}$CNRS, UMR 7095, Institut d'Astrophysique de Paris, F-75014, Paris, France\\
$^{9}$Sorbonne Universit\'es, UPMC Univ Paris 06, UMR 7095, Institut d'Astrophysique de Paris, F-75014, Paris, France\\
$^{10}$Kavli Institute for Particle Astrophysics \& Cosmology, P. O. Box 2450, Stanford University, Stanford, CA 94305, USA\\
$^{11}$SLAC National Accelerator Laboratory, Menlo Park, CA 94025, USA\\
$^{12}$Laborat\'orio Interinstitucional de e-Astronomia - LIneA, Rua Gal. Jos\'e Cristino 77, Rio de Janeiro, RJ - 20921-400, Brazil\\
$^{13}$Observat\'orio Nacional, Rua Gal. Jos\'e Cristino 77, Rio de Janeiro, RJ - 20921-400, Brazil\\
$^{14}$Department of Astronomy, University of Illinois, 1002 W. Green Street, Urbana, IL 61801, USA\\
$^{15}$National Center for Supercomputing Applications, 1205 West Clark St., Urbana, IL 61801, USA\\
$^{16}$Institut de Ci\`encies de l'Espai, IEEC-CSIC, Campus UAB, Facultat de Ci\`encies, Torre C5 par-2, 08193 Bellaterra, Barcelona, Spain\\
$^{17}$Institut de F\'{\i}sica d'Altes Energies (IFAE), The Barcelona Institute of Science and Technology, Campus UAB, 08193\\
$^{18}$George P. and Cynthia Woods Mitchell Institute for Fundamental Physics and Astronomy, and Department of Physics and Astronomy, Texas A\&M University, College Station, TX 77843,  USA\\
$^{19}$Department of Physics, IIT Hyderabad, Kandi, Telangana 502285, India\\
$^{20}$Department of Astronomy, University of Michigan, Ann Arbor, MI 48109, USA\\
$^{21}$Department of Physics, University of Michigan, Ann Arbor, MI 48109, USA\\
$^{22}$Institut de Ci\`encies de l'Espai, IEEC-CSIC, Campus UAB, Facultat de Ci\`encies, Torre C5 par-2, 08193 Bellaterra, Barcelona, Spain\\
$^{23}$Kavli Institute for Cosmological Physics, University of Chicago, Chicago, IL 60637, USA\\
$^{24}$Department of Astronomy, University of California, Berkeley,  501 Campbell Hall, Berkeley, CA 94720, USA\\
$^{25}$Lawrence Berkeley National Laboratory, 1 Cyclotron Road, Berkeley, CA 94720, USA\\
$^{26}$Kavli Institute for Particle Astrophysics \& Cosmology, P. O. Box 2450, Stanford University, Stanford, CA 94305, USA\\
$^{27}$Department of Astronomy, University of Illinois, 1002 W. Green Street, Urbana, IL 61801, USA\\
$^{28}$National Center for Supercomputing Applications, 1205 West Clark St., Urbana, IL 61801, USA\\
$^{29}$Astronomy Department, University of Washington, Box 351580, Seattle, WA 98195, USA\\
$^{30}$Cerro Tololo Inter-American Observatory, National Optical Astronomy Observatory, Casilla 603, La Serena, Chile\\
$^{31}$Australian Astronomical Observatory, North Ryde, NSW 2113, Australia\\
$^{32}$Departamento de F\'{\i}sica Matem\'atica,  Instituto de F\'{\i}sica, Universidade de S\~ao Paulo,  CP 66318, CEP\\
$^{33}$Department of Astrophysical Sciences, Princeton University, Peyton Hall, Princeton, NJ 08544, USA\\
$^{34}$Instituci\'o Catalana de Recerca i Estudis Avan\c{c}ats, E-08010 Barcelona, Spain\\
$^{35}$Jet Propulsion Laboratory, California Institute of Technology, 4800 Oak Grove Dr., Pasadena, CA 91109, USA\\
$^{36}$Department of Physics and Astronomy, Pevensey Building, University of Sussex, Brighton, BN1 9QH, UK\\
$^{37}$Centro de Investigaciones Energ\'eticas, Medioambientales y Tecnol\'ogicas (CIEMAT), Madrid, Spain\\
$^{38}$Universidade Federal do ABC, Centro de Ci\^encias Naturais e Humanas, Av. dos Estados, 5001, Santo Andr\'e, SP, Brazil, 09210-580\\
$^{39}$Computer Science and Mathematics Division, Oak Ridge National Laboratory, Oak Ridge, TN 37831\\
}

\appendix
\section{Colour Terms}
\label{CTerms}
The colour terms used in this analysis are included here for completeness.
\begin{align}
H_{\mathrm{VHS}} - H_{\mathrm{UKIDSS}} &= -0.01(i-z)_{\mathrm{SDSS}} + 0.02 \nonumber \\
i_{\mathrm{DES}} - i_{\mathrm{SDSS}} &= -0.30(i-z)_{\mathrm{SDSS}} + 0.02 \nonumber \\
J_{\mathrm{VHS}} - J_{\mathrm{UKIDSS}} &= -0.01(i-z)_{\mathrm{SDSS}} - 0.02 \nonumber \\
K_{\mathrm{VHS}} - K_{\mathrm{UKIDSS}} &= 0.04(i-z)_{\mathrm{SDSS}} - 0.07 \nonumber \\
Y_{\mathrm{DES}} - Y_{\mathrm{UKIDSS}} &= 0.09(i-z)_{\mathrm{SDSS}} - 0.08 \nonumber \\
z_{\mathrm{DES}} - z_{\mathrm{SDSS}} &= -0.07(i-z)_{\mathrm{SDSS}} - 0.01 \nonumber
\end{align}

Figures \ref{BDFit} and \ref{QFit} show the fits for the full range of models
used in this work for the highest ranked quasar and the highest ranked brown
dwarf.
\begin{figure*} 
\includegraphics[width = \linewidth]{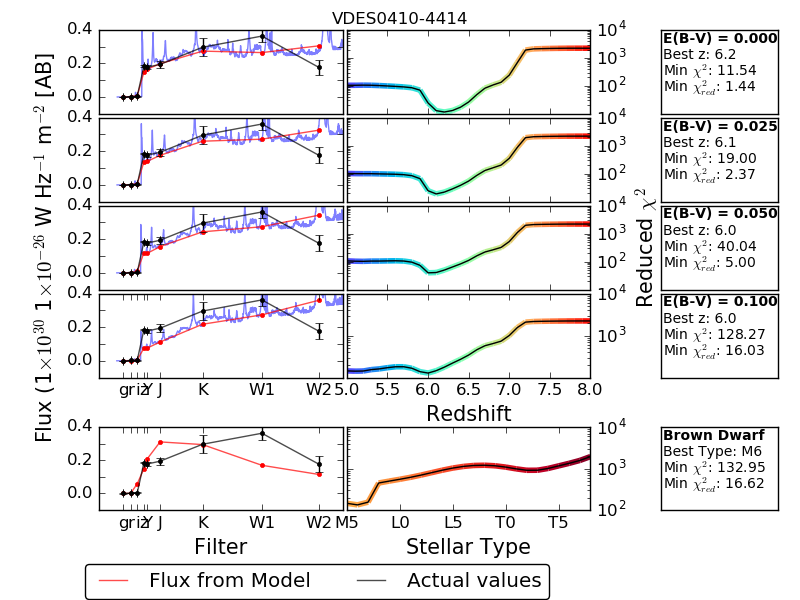}
\caption{An example of the fitting results for the highest ranked quasar in the
sample. The four different reddening models and the brown dwarf fit are shown
in the left column of plots and the right column shows the reduced $\chi^{2}$
fits for the range of redshifts/models considered. The brown dwarf model is
clearly different to any of the quasar models. Note the different scales on the
$\chi^{2}$ plots.} 
\label{QFit} 
\end{figure*}

\begin{figure*} 
\includegraphics[width = \linewidth]{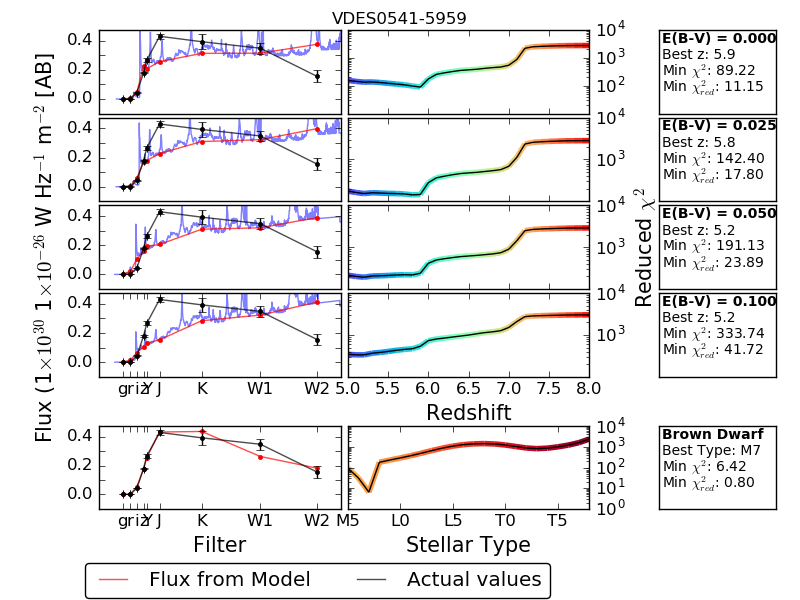}
\caption{An example of the fitting results for a probably brown dwarf found in the
sample. The four different reddening models and the brown dwarf fit are shown
in the left column of plots and the right column shows the reduced $\chi^{2}$
fits for the range of redshifts/models considered. It can be seen that the
brown dwarf model is closer to the data than any of the quasar models. Note the
different scales on the $\chi^{2}$ plots.} \label{BDFit} \end{figure*}

\end{document}